\newcommand{\mic}{$\mu$m}

\newcommand{\ltir}{$I_\text{TIR}$}

\newcommand{\hmol}{H$_2$}
\newcommand{\HI}{$\ion{H}{i}$}
\newcommand{\HII}{$\ion{H}{ii}$}
\newcommand{\CII}{$\left[\ion{C}{ii}\right]$}

\newcommand{\cplus}{C$^+$}

\documentclass{aa}  
\usepackage{tabularx}
\usepackage{caption}
\usepackage{subcaption}
\usepackage[switch, modulo]{lineno}
\usepackage{graphicx}
\usepackage{txfonts}
    \usepackage{multirow}
\usepackage[pdfpagelabels=false]{hyperref}	% Hyperlinks
\hypersetup{colorlinks=true,linkcolor=blue,citecolor=blue,filecolor=blue,urlcolor=blue}

\usepackage{orcidlink}

\newcommand{\eso}{European Southern Observatory (ESO), Karl-Schwarzschild-Stra{\ss}e 2, 85748 Garching, Germany}

\begin{document} 

   \title{Full disc \CII\ mapping of nearby star-forming galaxies}

  \subtitle{SOFIA FIFI/LS observations of NGC\,3627, NGC\,4321, and NGC\,6946} 

   \author{I.~Kova{\v{c}}i{\'{c}}\inst{1,2}\,\orcidlink{0000-0001-6751-3263}
            \and
            A.~T.~Barnes\inst{3,2}\,\orcidlink{0000-0003-0410-4504}
            \and
            F.~Bigiel\inst{2}\,\orcidlink{0000-0003-0166-9745}
            \and
            I.~De Looze\inst{1}\,\orcidlink{ 0000-0001-9419-6355}
            \and
            S.~C.~Madden\inst{4}\,\orcidlink{0000-0003-3229-2899}
            \and
            R.~Herrera-Camus\inst{5}\,\orcidlink{0000-0002-2775-0595}
            \and
             A.~Krabbe\inst{6}\,\orcidlink{0000-0002-8522-7006}
             \and
            M.~Baes\inst{1}\,\orcidlink{0000-0002-3930-2757}
          \and
            A.~Beck\inst{6}\,\orcidlink{0000-0002-1373-1377}
            \and
            A.~D.~Bolatto\inst{7}\,\orcidlink{0000-0002-5480-5686}
            \and
            A.~Bryant\inst{6}\,\orcidlink{0000-0002-9472-2391}
            \and
            S.~Colditz\inst{6}\,\orcidlink{0000-0002-5613-1953}        
            \and
            C.~Fischer\inst{6}\,\orcidlink{0000-0003-2649-3707}       
            \and
            N.~Geis\inst{8}            
            \and
            C.~Iserlohe\inst{6}\,\orcidlink{0000-0003-4223-7439}       
            \and
            R.~Klein\inst{9}\,\orcidlink{0000-0002-7187-9126}
            \and
            A.~Leroy\inst{10}\,\orcidlink{0000-0002-2545-1700}
            \and
            L.~W.~Looney\inst{11, 12}\,\orcidlink{0000-0002-4540-6587}          
            \and
            A.~Poglitsch\inst{8}\,\orcidlink{0000-0002-6414-9408}          
            \and
            N.~S.~Sartorio\inst{1}\,\orcidlink{0000-0003-2138-5192}
            \and
    W.~D.~Vacca\inst{13}\,\orcidlink{0000-0002-9123-0068} 
            \and
            S.~van~der~Giessen\inst{1, 14}\,\orcidlink{0009-0001-4144-9635}
            \and
            A.~Nersesian\inst{15,1}\,\orcidlink{0000-0001-6843-409X}
          }

   \institute{Sterrenkundig Observatorium, Universiteit Gent, Krijgslaan 281 / S9, 9000 Gent, Belgium 
         \and Argelander-Institut für Astronomie, Universität Bonn, Auf dem Hügel 71, 53121 Bonn, Germany
         \and \eso
         \and Université Paris-Saclay, Université Paris Cité, CEA, CNRS, AIM,
91191, Gif-sur-Yvette, France
        \and Departamento de Astronomía, Universidad de Concepción, Barrio Universitario, Concepción, Chile
        \and Deutsches SOFIA Institut, Universität Stuttgart, Pfaffenwaldring 29, D-70569 Stuttgart, Germany
        \and Department of Astronomy, University of Maryland, College Park, MD 20742, USA
        \and Max-Planck-Institut für extraterrestrische Physik, Gießenbachstrasse 1, D-85748 Garching, Germany
        \and SOFIA-USRA, NASA Ames Research Center, MS N232-12, Moffett Field, CA 94035-1000, USA
        \and Department of Astronomy, The Ohio State University, 140 West 18th Avenue, Columbus, OH 43210, USA
        \and Department of Astronomy, University of Illinois, 1002 W Green Street, Urbana, IL 61801, USA
        \and National Radio Astronomy Observatory, 520 Edgemont Rd. Charlottesville, VA 22903, USA
        \and NSF’s NOIRLab, 950 N. Cherry Avenue, Tucson, AZ 85719, USA
        \and Dept. Fisica Teorica y del Cosmos, Universidad de Granada, Granada, Spain
        \and STAR Institute, Université de Liège, Quartier Agora, Allée du Six Août 19c, 4000 Liège, Belgium 
        }

   \date{Received 2024 March 21; accepted 2024 November 28}

 \abstract{As a major cooling line of interstellar gas, the far-infrared 158 \mic\ line from singly ionised carbon \CII\ is an important tracer of various components of the interstellar medium in galaxies across all spatial and morphological scales. Yet, there is still not a strong constraint on the origins of \CII\ emission.}
 {In this work, we derive the resolved \CII\ star formation rate relation and aim to unravel the complexity of the origin of \CII.}
{We used the Field-Imaging Far-Infrared Line Spectrometer on board the Stratospheric Observatory for Infrared Astronomy to map \CII\ in three nearby star-forming galaxies at sub-kiloparsec scales, namely, NGC\,3627, NGC\,4321, and NGC\,6946, and we compared these \CII\ observations to the galactic properties derived from complementary data from the literature.}
{We find that the relationship between the \CII\ fine structure line and star formation rate shows variations between the galaxies as well as between different environments within each galaxy.}
{Our results show that the use of \CII\ as a tracer for star formation is much more tangled than has previously been suggested within the extragalactic literature, which typically focuses on small regions of galaxies and/or uses large-aperture sampling of many different physical environments. As found within resolved observations of the Milky Way, the picture obtained from \CII\ observations is complicated by its local interstellar medium conditions. Future studies will require a larger sample and additional observational tracers, obtained on spatial scales within galaxies, in order to accurately disentangle the origin of \CII\ and calibrate its use as a star formation tracer. }


   \maketitle
%
%-------------------------------------------------------------------
\section{Introduction}\label{sec:introduction}

\begin{table*}

\centering
    \caption{Properties of the galaxy sample.}
\label{tab:all_gal}
\resizebox{\textwidth}{!}{
\begin{tabular}{cccccccccccccc}
\hline\hline
Galaxy & RA & DEC & $i$ & PA & Morph. & Dist. & scale & $R_\mathrm{eff}$ & $V_\mathrm{LSR}$ & $M_\mathrm{H_I}$ & $M_\mathrm{H2}$ & $M_\mathrm{star}$ & SFR \\
 & $\mathrm{{}^{\circ}}$ & $\mathrm{{}^{\circ}}$ & $\mathrm{{}^{\circ}}$ & $\mathrm{{}^{\circ}}$ & & $\mathrm{Mpc}$ & $pc/"$ & $\mathrm{kpc}$ & km/s & log($\mathrm{M_{\odot}}$) & log($\mathrm{M_{\odot}}$) & log($\mathrm{M_{\odot}}$) & $\mathrm{M_{\odot}\,yr^{-1}}$ \\
     $(a)$ & $(b)$ & $(b)$ & $(c)$ &$(c)$ & $(d)$ & $(e)$ & $(f)$ & $(g)$ & $(h)$  & $(i)$ & $(j)$&$(k)$&$(l)$ \\ 
\hline
NGC~3627 & 170.063 & 12.991 & 57.3 & 173.1 & Sb & 11.3 & 55 & 3.6 & 715.4 & 9.1 & 9.8 & 10.8 & 3.8 \\
NGC~4321 & 185.729 & 15.822 & 38.5 & 156.2 & SABb & 15.2 & 70 & 5.5 & 1572.3 & 9.4 & 9.9 & 10.7 & 3.6 \\
NGC~6946 & 308.719 & 60.154 & 33.0 & 243.0 & SABc & 7.3 & 34 & 4.5 & 61.3 & 10.0 & 9.6 & 10.5 & 5.9 \\
\hline\hline
\end{tabular}}
    \tablefoot{
    $(a)$ Galaxy name. $(b)$ Central right ascension~(RA) and declination~(Dec) from \cite{Salo2015}.
    $(c)$ Position angle~(PA) from \cite{deBlok2008, Lang2020}.
    $(d)$ Morphological (Morph.) classification taken from HyperLEDA \citep{Makarov2014}.
    $(e)$ Source distances (Dist.) are taken from the compilation of \citet{Anand2021a, Anand2021b}.  
    $(f)$ Distance scale.
    $(g)$ Effective radius ($R_\mathrm{eff}$) that contains half of the stellar mass of the galaxy \citep{Leroy2021a}. 
    $(h)$ Centroid velocity ($V_\mathrm{LSR}$). Local standard of rest velocities taken from HyperLEDA \citep{Makarov2014} and \citet{Lang2020}.
    $(i)$ Total atomic gas mass ($M_\mathrm{HI}$) taken from HYPERLEDA \citep{Makarov2014}.
    $(j)$ Total molecular gas mass ($M_\mathrm{H2}$) from \citet{Leroy2013, Leroy2021a}. 
    $(k)$ Global star formation rate (SFR) derived by \citet{Leroy2021a}, using \textit{GALEX}~UV and \textit{WISE}~IR photometry, following a similar methodology to \cite{Leroy2019}.
    }

\end{table*}

Singly ionised carbon (\cplus) exists in various gas phases, including neutral gas \citep{Wolfire_2003}, photodissociation regions \citep[PDRs;][]{Crawford,Tielens_1985,Hollenbach_1991,Bakes_1994,Bakes_1998, Malhotra, Boselli_2002,Pierini_2003}, and ionised gas \citep{Wolfire_1995,Nakagawa_1998}. This is due to both the high abundance of carbon and its low ionisation potential \citep[$11.3 {\rm \, eV}$;][]{Stacey_1991,stacey_2010}. In addition, due to its low excitation energy ($E/k \sim 92 {\rm \, K}$), \cplus\ is easily excited in both neutral and ionised gas, where the \CII\ forbidden transition can become a major cooling line \citep{Wolfire_2003}. Due to these properties, the \CII\ 158\,\mic\ emission line plays an important role in the interplay between star formation regions and the interstellar medium (ISM). 

The \CII\ emission is often used to estimate the star formation rate \citep[SFR;][]{Pierini_1999, Leech_1999, De_Looze_2011, de_looze_2014, Herrera, Pineda_2018}, especially in high-$z$ galaxies \citep{stacey_2010,Lagache_2018}. However, there are issues limiting its usage, such as \CII\ deficit: the lower \CII\ to total infrared (TIR) luminosity ratio observed for warmer sources, such as the ultraluminous infrared galaxies (ULIRGs), as measured by their high infrared (IR) colour \citep{Malhotra_1997,Garcia_Carpio_2011,Herrera,Smith_2017,Diaz_Santos_2017}. There are several proposed explanations for the \CII\ deficit. For example, in areas of high UV radiation, dust grains become positively charged, which results in a higher Coulomb barrier for the photoelectrons to overcome \citep{Croxall_2012}. A compact IR source in the centre of the galaxy, such as an obscured active galactic nucleus (AGN), can lead to a lower \CII/TIR ratio by reducing the relative cooling effect of \CII\ by, for example, photodestruction of small grains by X-rays \citep{Malhotra_1997,Langer_2015}. The \CII\ line may be optically thick due to self-absorption by lower excitation foreground \cplus gas \citep{Russell_1980}. Another contribution could come from the dominance of other cooling channels. At high temperatures ($T \gg 100 {\rm \,K}$), the $\left[\ion{O}{i}\right]$\ $63{\rm \,\mu m}$\ emission becomes dominant, while at lower temperatures ($T < 92 {\rm \, K}$, below the \CII\ $158 {\rm \,\mu m}$ excitation temperature), \cplus\ is neutralised by neutral or negative polycyclic aromatic hydrocarbon (PAH) molecules, causing the CO emission to become dominant \citep{Kaufman_1999}.

For high redshift galaxies, \CII\ emission is more extended than UV emission, which causes the \CII\ surface brightness to be lower than expected from local environments \citep{Ferrara_2019, LeFevre_2020}. Due to the compact sizes of high-$z$ galaxies, intense radiation fields are also more effective at performing photoevaporation of molecular clouds, which suppresses the peak in the \CII\ spectrum and implies a modified Kennicutt-Schmidt relation \citep{Vallini_2015}.

The {\it Herschel} Space Observatory \citep[e.g.][]{Pilbratt_2010} has provided some of the first insights into the IR and sub-millimetre wavelengths. While its sensitivity was unrivalled, it was limited in mapping area for spectroscopic observations. Where observations of \CII\ across galactic discs are available, they are often limited to narrow strips across the galaxies. Hence, the motivation to further study the \CII\ fine structure line throughout the nearby galaxies remains strong. Additionally, at high redshifts ($z\geq1$), the \CII\ 158\,\mic\ line becomes observable from the ground (\citealp{Maiolino}, \citealp{Walter}, originally proposed by \citealp{Petrosian}). This line is a good tracer of the bulk of the molecular gas reservoir independent of metallicity \citep{Ramambason_2024} and thus plays a particularly important role in low metallicity environments, where CO is difficult to observe and \CII\ is relatively bright \citep[][]{Madden_2020}. It should also remain a good tracer at extreme redshifts \citep[up to $z\approx 20$;][]{de_Blok}. 

The aim of this study is to evaluate the applicability of \CII\ as a tracer for the SFR within the different environments of a galaxy. In order to perform this analysis, we used full disc observations of three nearby actively star forming disc galaxies by the Field-Imaging Far-Infrared Line Spectrometer \citep[FIFI-LS;][]{FIFI} from the Stratospheric Observatory for Infrared Astronomy \citep[SOFIA;][]{SOFIA, Young_2012}.

This work is organised as follows: In Section \ref{sec:data} we describe the new SOFIA/FIFI-LS \CII\ observations and ancillary data. Section \ref{sec:results} shows the spatial distribution and average radial trends for \CII\ and ancillary data. In section \ref{sec:analysis} we analyse the performance of \CII\ as a diagnostic for various states of the ISM. Section \ref{sec:summary} summarises our findings.

\section{Observations}\label{sec:data}

\subsection{Target selection}

We selected the three star-forming galaxies: NGC\,3627, NGC\,4321, and\,NGC\,6946. Due to their proximity (7.3 to 15.2\,Mpc), the SOFIA observations profit from the high spatial resolution ($\sim 0.5-1.5 {\rm \, kpc}$) sampling a variety of local environmental conditions. Additionally, these galaxies are well studied at many other wavelengths, bringing the advantage of available complementary data for the analysis. A summary of properties of these galaxies can be found in Table \ref{tab:all_gal}, and all the complementary data used in this work is described in Table \ref{tab:beams}. Here, we briefly introduce the main characteristics of the galaxies in our sample, including the nearby galaxy NGC\,6946, for which full \CII\ disc mapping was presented in \citet{Bigiel_2020}.

\begin{itemize}
    \item NGC\,3627 is a barred spiral galaxy that contains an AGN \citep{Moustakas_2010} and resides at a distance of 11.3\,Mpc. Along with NGC\,3623 and NGC\,3628 \citep{garcia_1993}, it is part of a close group of galaxies known as the Leo triplet. Due to a past gravitational interaction with NGC\,3628, the galaxy shows an accumulation of gas mass in the spiral arms \citep{zhang_1993}.
    \item NGC\,4321, also known as M100, is an intermediate spiral galaxy that contains an AGN \citep{Moustakas_2010} and resides at a distance of 15.2\,Mpc. It is one of the brightest galaxies in the Virgo Cluster \citep{Binggeli}, and as is common for the members this cluster, it is deficient in hydrogen throughout the disc, especially in the centre \citep{Chung_2010}, which has an effect on its star formation \citep{Koopmann}.
    \item NGC\,6946, also known as the Fireworks Galaxy due to its high number of supernovae \citep{Tran_2023}, is a double-barred \citep{Schinnerer_2006} intermediate spiral galaxy with a high SFR throughout the disc \citep{Sauty} residing at a distance of 7.3\,Mpc. It is an isolated galaxy belonging to the Local Void with an extended atomic hydrogen disc \citep{Peebles}. Throughout the disc, the galaxy has several atomic hydrogen pockets, some expanding out to a radius of 2\,kpc \citep{Efremov_2016}.

\end{itemize}

\begin{table}[ht]
        \caption{Summary of the observational data used in this work.}
\centering
\resizebox{\columnwidth}{!}{
    \begin{tabular}{l c c}
    \hline\hline
        Band & $\lambda$ [\mic ]  & PSF FWHM [\arcsec]\\
        \hline
      \small{$\text{$\left(a\right)$}$ SOFIA FIFI-LS \CII}  &  158 &  $15.6$ \\
       \small{$\text{$\left(b\right)$}$ HERACLES CO(2-1)} &  1300 & $13$ \\
       \small{$\text{$\left(c\right)$}$ VLA THINGS \HI } & 21 106 & $7$\\
       \small{$\text{$\left(d\right)$}$ VLA VIVA \HI} & 21 106  & $15$\\
       \small{$\text{$\left(e\right)$}$ WISE} & 3.4, 4.6, 12, 22 & $6.1, 6.4, 6.5, 12$ \\
        \small{$\text{$\left(f\right)$}$ GALEX FUV} & 0.154 & 4.5 \\
        \small{$\text{$\left(g\right)$}$ \textit{Herschel} PACS} & 70, 100, 160 &  5.6, 6.8, 11.3\\
        \small{$\text{$\left(h\right)$}$ \textit{Spitzer} IRAC} & 3.6, 4.5, 5.8, 8 & $1.2$ \\
        \small{$\text{$\left(i\right)$}$ \textit{Spitzer} MIPS} & 24& $6$ \\
        \footnotesize{$\text{$\left(j\right)$}$ Dust/starlight models} & & $18.2$ \\
        \footnotesize{$\text{$\left(k\right)$}$ Wise and GALEX Atlas} & & 15 \\
        \hline \hline
    \end{tabular}
    }
        \tablefoot{ 
        $(a)$ \citet{cii_psf}.
        $(b)$ \citet{Leroy_2009}.
        $(c)$ \citet{Walter_2008}.
        $(d)$ \citet{Chung_2010}.
        $(e)$ \citet{wise}.
        $(f)$ \citet{galex}.
        $(g)$ \citet{Poglitsch_2010}.
        $(h)$ \citet{IRAC}.
        $(i)$ \citet{MIPS}.
        $(j)$ \citet{Aniano_2020}.
        $(k)$ \citet{Leroy_2019}.}
    \label{tab:beams}
\end{table}

\subsection{SOFIA observations}\label{sec:sofia_obs}

\begin{figure*}[h!]
    \centering
    \includegraphics[width=\textwidth]{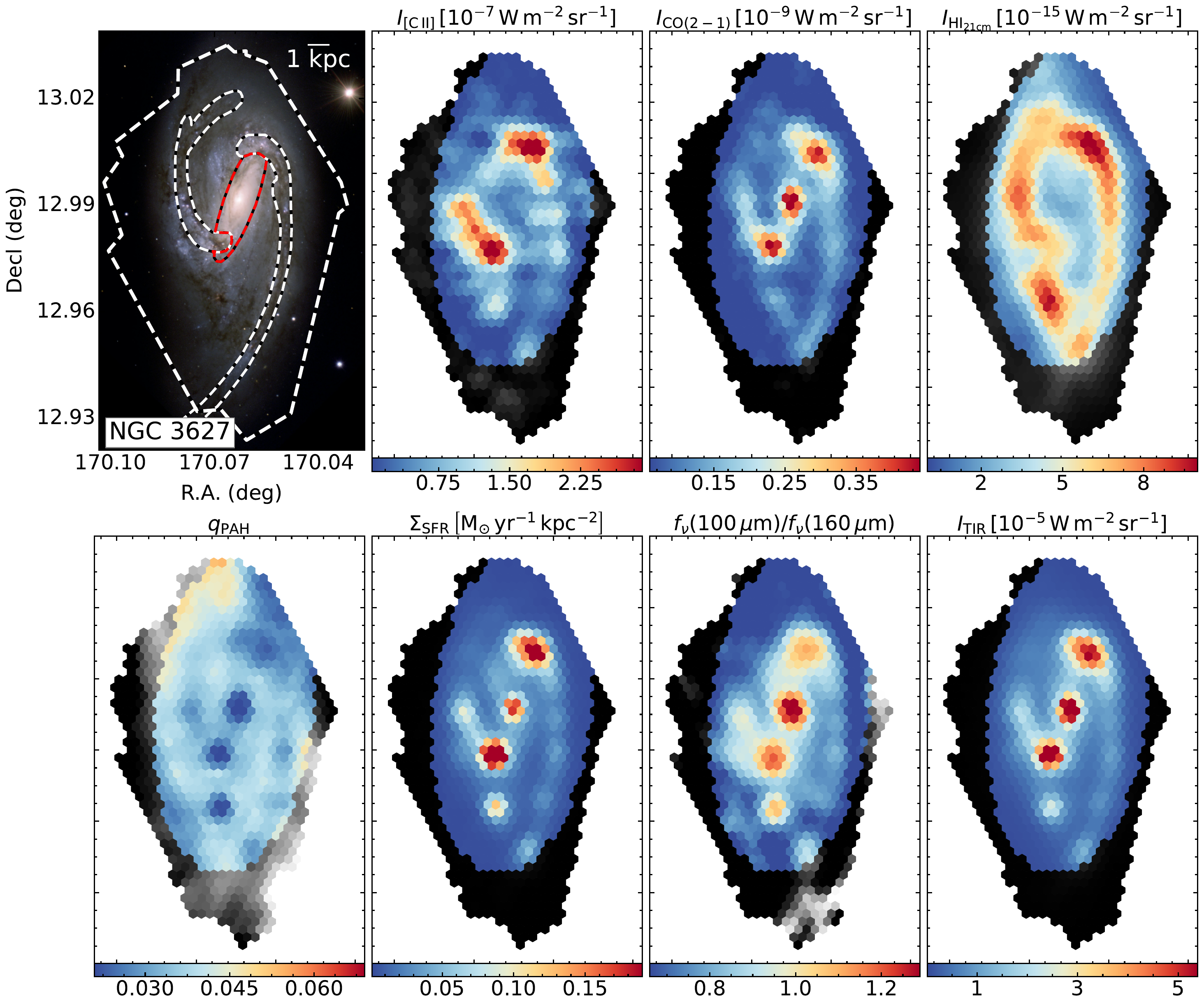}
        \caption{Full disc maps for galaxy NGC\,3627. \emph{Top:} RGB (image credit: ESO/Pieter Barthel), \CII, CO(2-1), and \HI\ 21-cm integrated intensity maps. \emph{Bottom:} $q_\text{PAH}$, $\Sigma_\text{SFR}$, $f_{\nu} (100 {\rm \, \mu m}) / f_{\nu} (160 {\rm \, \mu m})$ ratio as a tracer of dust temperature and TIR integrated intensity maps for galaxy NGC\,3627, all convolved to the resolution of 18.2\arcsec. The area in the grey colour scale marks the region outside the maximum considered galactocentric radius ($r = 9 {\rm\,kpc}$) that was not used in the analysis. In the RGB image, the outer white dashed contour outlines the extent of the SOFIA/FIFI-LS observations. Three distinct regions, centre (which consists of the galactic centre and bar), arm, and interarm, are respectively outlined in red and white contours.
        }
\label{fig:gal_struct}

\end{figure*}

\begin{figure*}[ht]
    \includegraphics[width=0.99\textwidth]{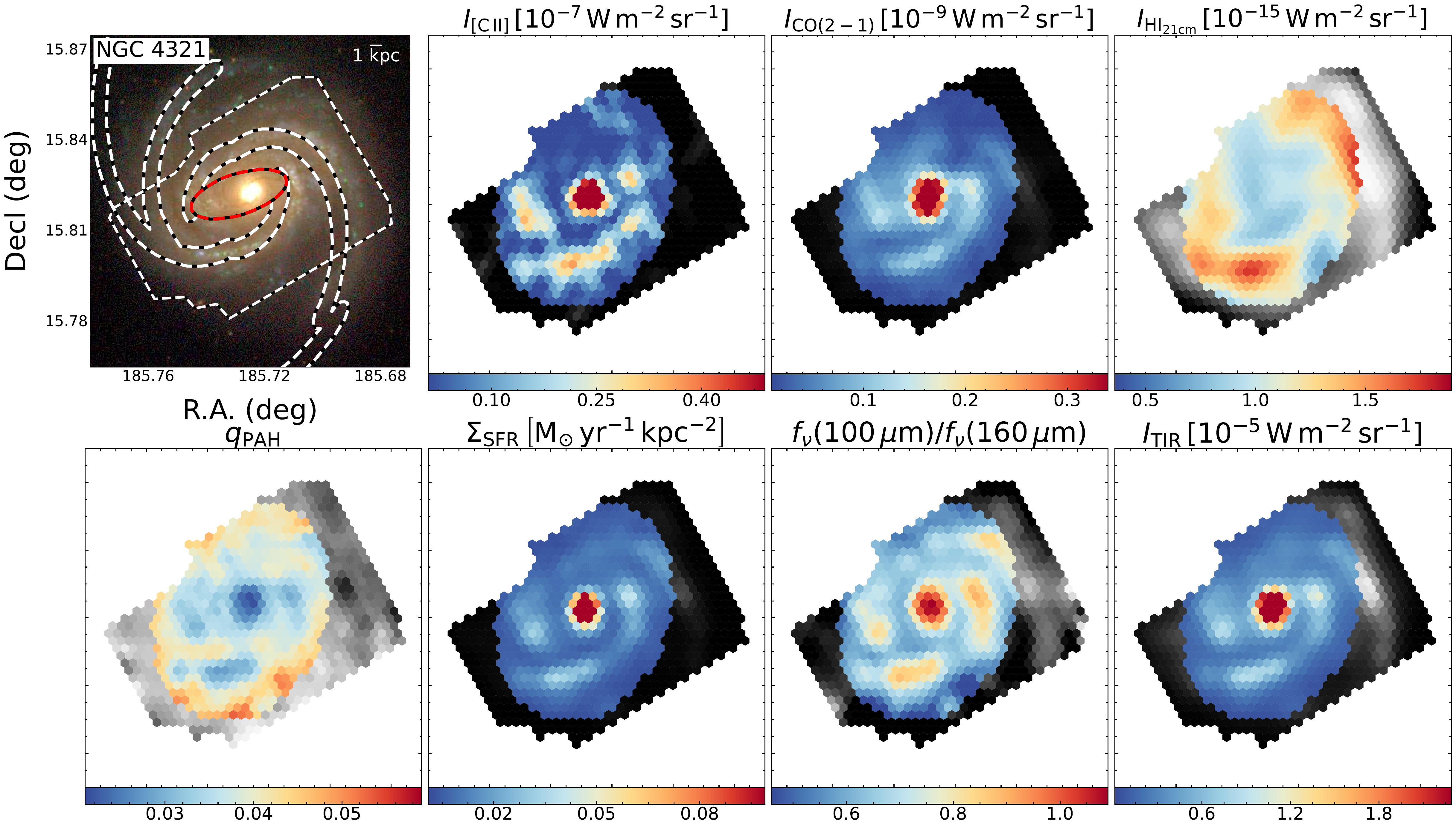}
        \caption{Similar to Fig.\,\ref{fig:gal_struct} but for NGC\,4321. The maximum considered galactocentric radius is $r = 9 {\rm \, kpc}$. RGB image credit: SDSS9.}
    \label{fig:gal_struct4321}

\end{figure*}
\begin{figure*}[ht]
    \includegraphics[width=\textwidth]{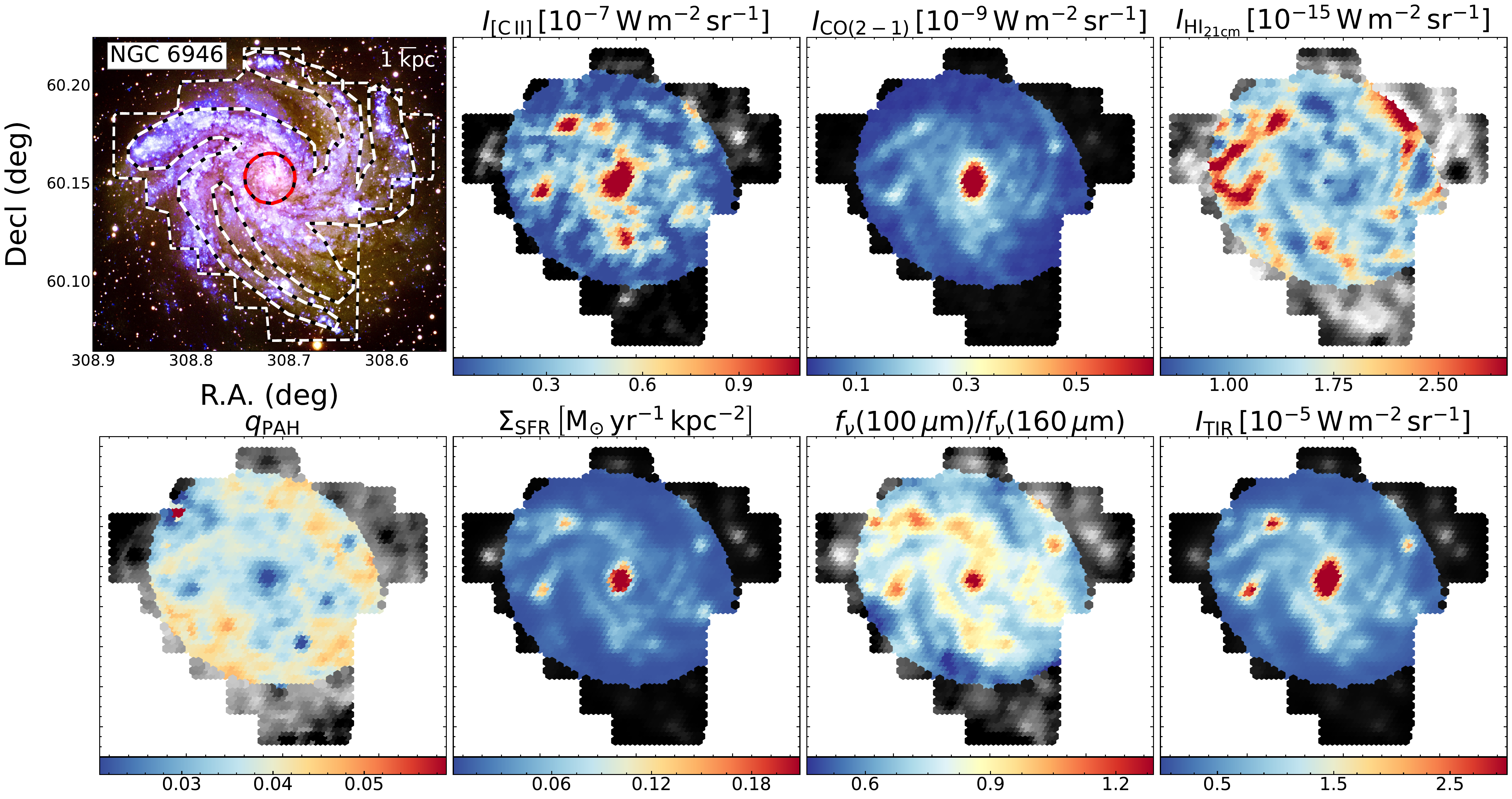}

    \caption{Similar to Fig.\,\ref{fig:gal_struct} but for NGC\,6946. The maximum considered galactocentric radius is $r = 8 {\rm \, kpc}$. RGB image credit: V. Testa, C. DeSantis, LBTO.}
    \label{fig:gal_struct6946}
\end{figure*}

SOFIA consisted of a 2.5\,m telescope carried by a Boeing 747 aircraft operating at altitudes above most of the atmospheric water vapour and covered the IR region of the spectrum. Its instrument FIFI-LS featured two parallel spectral channels with wavelengths of $51-125{\rm\,\mu m}$, the `blue channel', and $115-203 {\rm \,\mu m}$, the `red channel'. Each channel consisted of a field of view (FOV) of $5 \times 5$ spatial pixels, with a plate scale of $6.14\arcsec \times 6.25\arcsec$ for blue and $12.2\arcsec \times 12.5\arcsec$ for red. This translates into an FOV of $60\arcsec \times 60\arcsec$ for the red channel and $30\arcsec \times 30\arcsec$ for the blue channel \citep{Colditz_2018,Fischer_2018}. The spatial resolution at the observed \CII\ 158\,$\mu{\rm m}$ line is 15.6\arcsec\ \citep{cii_psf}, with a spectral resolution of $R = 1200$.

Data were processed as described in the FIFI-LS data reduction pipeline \citep{pipeline} and was corrected for atmospheric absorption with the transmission curves from ATRAN \citep{atran}. Baseline subtraction was performed after visually inspecting the spectra and fitting a first-order polynomial to the outside region on both sides of the signal-containing range ($-400 {\rm \, km \, s^{-1}}$ to $400 {\rm \, km \, s^{-1}}$) and subtracting the fit. After the baseline subtraction, the SOFIA/FIFI-LS \CII\ 158\,\mic\ fine structure line integrated intensity maps as well as the complementary data (Section \ref{sec:compl_data}) were projected to a common resolution of 18.2\arcsec\ using the \texttt{PyStructure} code,\footnote{https://github.com/jdenbrok/AG\_Bigiel} and oversampled by half a 18.2\arcsec\ beam on a hexagonal grid, ensuring that the pixel coverage better matches the beam. This corresponds to spatial resolutions of 0.50\,kpc for NGC\,3627, 0.67\,kpc for NGC\,4321, and 0.32\,kpc for NGC\,6946.
The SOFIA maps were cross-calibrated with the Herschel PACS \CII\ maps from the KINGFISH program \citep{Kennicutt_2011} and found to be consistent (see Appendix \ref{appendix-crossc}).

\subsection{Complementary data}\label{sec:compl_data}

We used complementary data from the literature to map out other properties for comparison with our \CII\ observations. These include atomic gas surface density ($\Sigma_\text{atom}$), molecular gas surface density ($\Sigma_\text{mol}$), SFR surface density ($\Sigma_\text{SFR}$), total infrared and sub-millimetre energy budget ($I_\text{TIR}$), and PAH intensity ($I_\text{PAH}$).

The \HI\ (21\,cm) line observations were obtained from the \HI\ Nearby Galaxy Survey \citep[THINGS\footnote{https://www2.mpia-hd.mpg.de/THINGS/Data.html};][]{Walter_2008} for NGC\,3627 and NGC\,6946, and VLA Imaging of Virgo in Atomic Gas \citep[VIVA\footnote{http://www.astro.yale.edu/viva/};][]{Chung_2010} for NGC\,4321, and they were used to model the $\Sigma_\text{atom}$, using Eq. 4 from \citet{Jim_nez_Donaire_2019}, with the factor 1.36 for helium not included:
\begin{equation}
    \Sigma_\text{atom} {\rm \, \left[ M_{\odot}\, pc^{-2}\right]} 
    = 0.020 \cdot I_\text{21\,cm}{\rm \, \left[ K\,km\, s^{-1}\right]}\cdot \cos i\quad,
\end{equation}
with $\cos i$ being the inclination correction (Table \ref{tab:all_gal}).

 The $^{12}$CO $J = 2 \rightarrow 1$, henceforth CO(2-1), was obtained from the HERA CO-Line Extragalactic Survey \citep[HERACLES\footnote{https://iram-institute.org/science-portal/proposals/lp/completed/lp001-the-hera-co-line-extragalactic-survey/};][]{Leroy_2009} and used to model $\Sigma_\text{mol}$ as described in Eq. 2 from \citet{Sandstrom_2013}:
 \begin{equation}
    \Sigma_\text{mol} {\rm \,\left[M_{\odot}\, pc^{-2}\right]} = \alpha_\text{CO} I_\text{CO(2-1)}{\rm\,\left[K\, km\, s^{-1}\right]} \cdot \cos i \quad.
\label{eq:alpha_co}\end{equation}
They derived the conversion factor from CO to total \hmol\ mass,
 $\mathrm{\alpha_{CO}}$, directly from HERACLES CO(2-1) \citep{Leroy_2009} and THINGS \HI\ \citep{Walter_2008}, modelling it as a function of galactocentric radius to take into account the so-called CO-dark molecular gas, which can be more prominent in the outer parts of galaxies with lower metallicities \citep{Wolfire_2010,Madden_2020}. It is important to note that the CO(2-1) map for NGC\,3627 suffers from calibration issues \citep{Leroy_2021}. We did have access to the new ALMA arcsecond CO(2-1) images \citep{Leroy_2021}; however they do not have the required coverage for NGC\,3627. We compared the ALMA and HERACLES CO(2-1) images, and while there are differences, they are fairly minor ($0.06^{+0.11}_{-0.26} {\rm \, dex}$). Regardless, this adds a layer of uncertainty that should be taken into consideration, specifically for the CO(2-1) full disc map (Fig.\,\ref{fig:gal_struct}), CO(2-1) and $\Sigma_\text{mol}$ radial profiles (Fig.\,\ref{fig:all_three_binning_combined}), and the $M_\text{mol}$ and \CII/CO(1-0) ratio in Sect.\,\ref{sec:mol_gas_tracer}, since $\Sigma_\text{mol}$ and CO(1-0) were modelled from HERACLES CO(2-1).

The $^{12}$CO $J = 1 \rightarrow 0$, henceforth CO(1-0), was modelled from the CO(2-1), assuming the CO(2-1)/CO(1-0) line ratio as described in Fig. 3 of \citet{Den_Brok_2021}, because the available CO(1-0) maps for these galaxies are of a much lower resolution. They used EMPIRE CO(1-0) \citep{Jim_nez_Donaire_2019} and HERACLES CO(2-1) maps and derived the resulting CO(2-1)/(1-0) radius-dependent line ratio values. Due to its very small variations with radius, the value for NGC\,6946 was taken as a constant 0.66.

The values for the $\alpha_\text{CO}$ factor used here were determined by \citet{Sandstrom_2013} using the CO(2-1) transition and thus include the fact that the conversion factor varies radially in the galaxies. In fact, wide variations can be seen spatially for all of these galaxies, with one of the lowest values determined for NGC\,3627 (see their Fig.\,7). For this reason, we did not simply apply the Milky Way $\alpha_\text{CO} = 4.4{\rm \, M_{\odot}\, pc^{-2} \, \left( K \, km \,s^{-1}\right)^{-1}}$ throughout our study.

The $\Sigma_\text{SFR}$ was estimated using IR and UV images by NASA's Wide-field Infrared Survey Explorer \citep[WISE;][]{wise} and the Galaxy Evolution Explorer \citep[GALEX;][]{galex}. These maps are available as a part of the PHANGS sample \citep{Leroy_2019}.

We used the recipe from \citet{Galametz_2013} to estimate \ltir\ with \textit{Herschel} PACS\footnote{https://www.cosmos.esa.int/web/herschel/pacs-point-source-catalogue} \citep{Poglitsch_2010} observations, which are available as a part of the Key Insights on Nearby Galaxies: a Far-Infrared Survey with the \textit{Herschel} \citep[KINGFISH\footnote{https://www.ipac.caltech.edu/publication/2011PASP..123.1347K};][]{Kennicutt_2011} galaxy sample. We modelled \ltir\ using only $70{\rm \, \mu m}$, $100{\rm \, \mu m}$, and $160{\rm \, \mu m}$, as the inclusion of longer wavelengths does not seem to have a severe impact \citep{Aniano_2020}.
Regarding $I_\text{PAH}$, it was modelled with the \textit{Spitzer} {\sc IRAC}\footnote{https://www.cfa.harvard.edu/irac/} 8\,\mic\ band, correcting for stellar emission with the IRAC 3.6\,\mic\ band, following Eq. 2 from \citet{Croxall_2012}:
 \begin{equation}
     I_\text{PAH} = \left[\nu S_{\nu}\left(8.0\right) - 0.24 \cdot \nu S_{\nu} \left(3.6\right)\right]\quad,
 \end{equation}
where $\nu$ denotes the frequency of the IRAC band and $S_{\nu}$ denotes the surface brightness at the indicated band.

Since we wished to compare the behaviour of \CII\ across different environments within the galaxies and from galaxy to galaxy, we defined different regions of interest: the galaxy centre (which consists of the galactic bar and central region for NGC\,3627 and NGC\,4321), the arm, and the interarm regions, based on 2D photometric decomposition of the \textit{Spitzer} IRAC $3.6 {\rm \, \mu m}$ images as part of the \textit{Spitzer} Survey of Stellar Structure in Galaxies (S$^4$G). The size and orientation of bars were identified visually, and spiral arms were included, as they strongly dominate the galactic disc. A log-spiral function was fitted to the bright regions along arms on the NIR images and assigned a width determined empirically based on CO emission \citep{salo2015spitzer, querejeta_2020}. For NGC\,6946, we traced the spiral arms in the PACS 70\,\mic\ band, as it showed clearer features than 3.6\,\mic\ used in S$^4$G. The central region for NGC\,6946 was determined as its inner 1.5\,kpc as described in \citet{Bigiel_2020}.

To analyse the relationship between the \CII\ and the interstellar dust, we used the dust mass fraction contributed by polycyclic aromatic hydrocarbons ($q_\text{PAH}$) from the KINGFISH galaxy sample dust and starlight models\footnote{ http://arks.princeton.edu/ark:/88435/dsp01hx11xj13h} \citep{Aniano_2020}. The authors used the \citet{DL07} dust model with `Milky Way' grain size distribution \citep{Weingartner_2001} and the radiation field heating the dust as estimated by \citet{Mathis_1983}.

\section{The morphology and multi-phase distribution of \CII\ emission}\label{sec:results}

To understand in which gas phases the \CII\ emission originates, we discuss the distribution of the \CII\ emitting regions and how it compares with tracers of the atomic and molecular gas. First, our analysis is on resolved scales (Section \ref{sec:cii_emiss}), and then we look at the averaged radial profiles (Section \ref{sec:rad_prof}).

\subsection{Spatial distribution of \CII\ emission}\label{sec:cii_emiss}

The \CII\ integrated intensity maps of NGC\,3627, NGC\,4321, and NGC\,6946 are presented in Figs.\,\ref{fig:gal_struct}, \ref{fig:gal_struct4321}, and \ref{fig:gal_struct6946}, along with the CO(2-1) and \HI\ $21 {\rm \, cm}$ maps. In NGC\,4321 and NGC\,6946, the \CII\ peaks towards the nuclei and presents emission peaks along the spiral arms, corresponding overall to the structures seen in the CO(2-1) map. NGC\,3627, however, shows a strong central \CII\ deficit, with an accumulation of \CII\ at the ends of the bar. CO(2-1), on the other hand, peaks at the centre but also follows the strong emission in the inner spiral arms. Bright \CII\ emission is patchy along the spiral arms, while the CO(2-1) prominently traces them. The \HI\ does not strictly follow the structures seen in \CII\ or CO(2-1) but is generally bright along the spiral arms. However, its structure is patchier for NGC\,6946. As is typically seen in spiral galaxies, the \HI\ distribution shows a void in the centre of each of the galaxies.

\subsection{Radial profiles}\label{sec:rad_prof}

\begin{figure*}
    \centering
    \includegraphics[width=\textwidth]{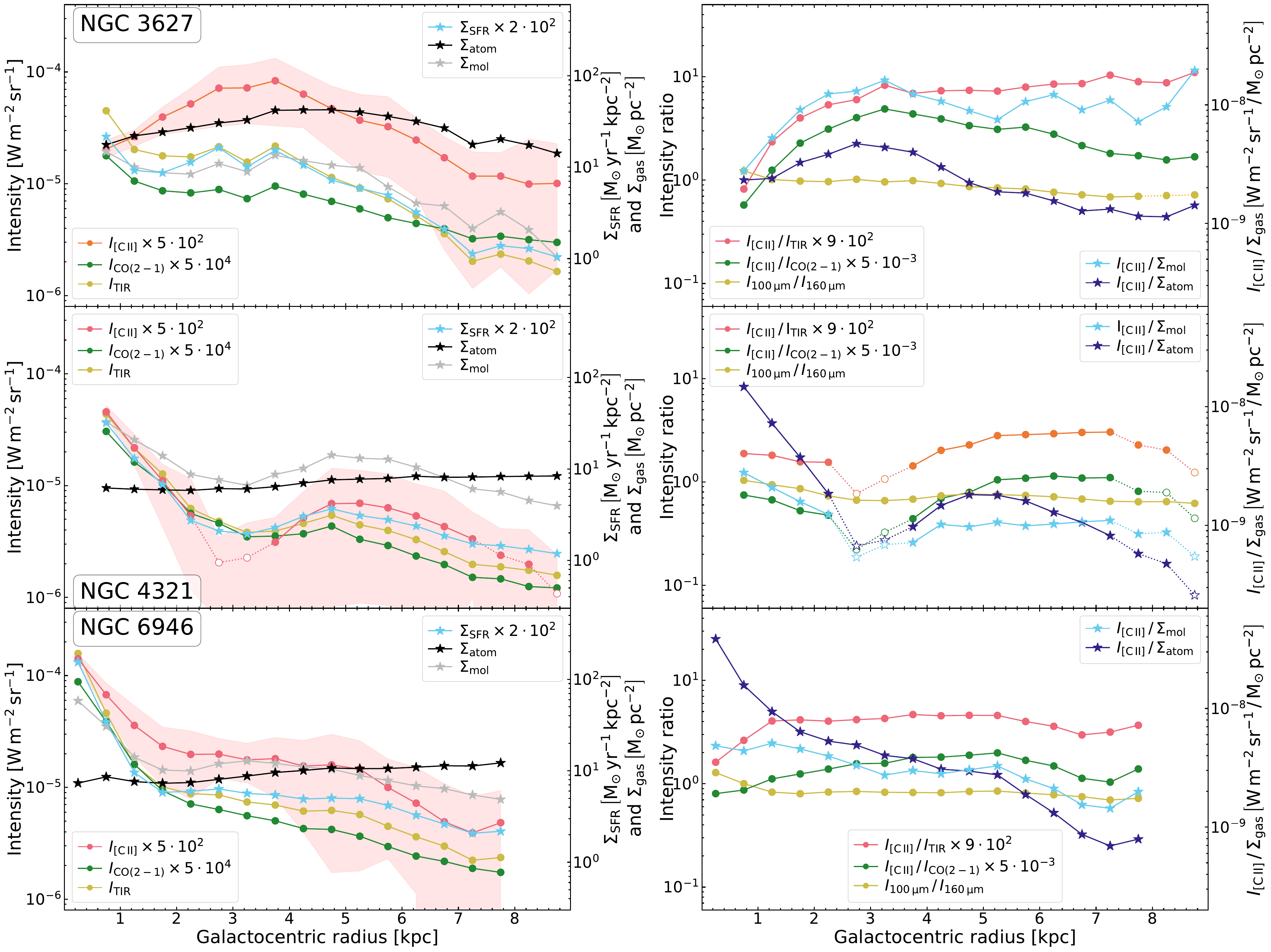}

    \caption{Radial profiles for galaxies NGC\,3627 (top), NGC\,4321 (middle), and NGC\,6946 (bottom).\emph{Left}: \CII, CO(2-1) and TIR intensities as well as surface densities of atomic and molecular gas and the SFR. \emph{Right}: Intensity ratios for \CII/TIR, \CII/CO(2-1), and $f_{\nu}(100{\rm \, \mu m})/f_{\nu} (160 {\rm \, \mu m})$ as well as ratios of \CII\ intensity and atomic or molecular gas surface density. Solid symbols show values with greater than $3\sigma$ significance, while hollow symbols show values with less than $3\sigma$ significance for \CII. The shaded area represents $\pm 1\sigma$ scatter for \CII\ in each radial bin.}
    \label{fig:all_three_binning_combined}
\end{figure*}

In this section, we describe the radial trends for all three galaxies (Fig.\,\ref{fig:all_three_binning_combined}), comparing the behaviour of \CII, as a function of galactocentric radius, relative to other tracers, such as the atomic and molecular gas and TIR.
As seen in the full disc maps (Fig.\,\ref{fig:gal_struct}), the \CII\ in NGC\,3627 has a drop in the centre, where CO(2-1) and $I_\text{TIR}$ are at their maximum intensities. The \CII\ intensity rises from the centre to its peak value between a $2-3 \text{\,kpc}$ radius where the ends of the bars and spiral arms shine brightly in the \CII. The $\Sigma_\mathrm{atom}$ radial profile rises from a dip in the centre to a broad peak near $4 {\rm \,kpc}$. CO(2-1), TIR, and $\Sigma_\mathrm{mol}$ show central peaks followed by a secondary smaller peak at $r \approx 3 {\rm \, kpc}$. Above this radius, \CII\ and TIR intensities as well as molecular gas surface density and SFR fall steadily, while CO(2-1) shows a milder decrease. At $r \approx 7 {\rm\, kpc}$, the \CII\ radial profile reverses mildly, and so does CO(2-1) to an extent. All the other values continue to drop until the radius cut-off at 9\,kpc. Intensity ratios as well as the $I_\text{\CII} / \Sigma_\mathrm{gas}$ ratio show a very similar rising structure in the inner 3\,kpc of the disc; however, after the peak, they start to diverge, with \CII/TIR remaining relatively constant, while CO(2-1) and gas ratios decrease until a $r\approx 7{\rm\,kpc}$. At this point, \CII/TIR, $I_\text{\CII}/\Sigma_\mathrm{atom}$, and $I_\text{\CII}/\Sigma_\mathrm{mol}$ show a reversal, but \CII/CO(2-1) evens out. The radial profile of $f_{\nu}(100\,\mu \mathrm{m}) / f_{\nu}(160\,\mu \mathrm{m})$, a tracer of dust temperature \citep{Smith_2019}, is flat, except for a slight central increase for all three galaxies.

The \CII\ data of NGC\,4321, the most distant of the three galaxies, has a low signal-to-noise ratio, resulting in a large number of points that fall below the $ 3 \sigma$ significance threshold, which are marked with open circles (Fig.\,\ref{fig:all_three_binning_combined}, middle row). The \CII, CO(2-1), and TIR intensities as well as the $\Sigma_\text{mol}$ and $\Sigma_\text{SFR}$ all peak at the centre, followed by a steady decline until $r \approx 3 {\rm \,kpc}$. This is followed by a smaller secondary peak at $r \approx 5{\rm  \, kpc}$, where the spiral arms exit, then a slow decline until the radius cut-off at 9\,kpc. Throughout the disc, $\Sigma_\mathrm{atom}$ is almost flat and seemingly not correlated with the behaviour of \CII\ or CO(2-1). In the full disc map (Fig.\,\ref{fig:gal_struct4321}), the \HI\ intensity, a tracer of $\Sigma_\mathrm{atom}$, shows an indication of the spiral arms; however, it extends into the interarm region on both sides. The \CII/TIR ratio follows \CII/CO(2-1) consistently throughout the disc, but it follows $I_\text{\CII}/\Sigma_\text{mol}$ less. The radial profile of $I_\text{\CII}/\Sigma_\text{atom}$ shows the highest deviation.

For NGC\,6946, the central peak of \CII, CO(2-1), TIR, $\Sigma_\text{mol}$, and $\Sigma_\text{SFR}$ up to $r \approx 1.5{\rm \, kpc}$ was chosen as the boundary between the galactic centre and disc. At $r \geq 1.5 {\rm \, kpc}$, the profiles even out. \CII\ shows a prominent drop after a small bump created from crossing a spiral arm at $r \approx 5 {\rm \, kpc}$. This behaviour, albeit a little less prominent, is visible in other tracers as well, and their slight deviation from \CII\ is visible in the ratios of the tracers. $\Sigma_\mathrm{atom}$ has a small bump in the centre at $r\approx 1.5 {\rm \,kpc}$ followed by a steady rise throughout the rest of the galaxy. \CII/TIR and \CII/CO(2-1) follow each other relatively well throughout the entire radial profile. In contrast, $I_\text{\CII}/\Sigma_\text{mol}$ drops noticeably from $r \approx 1.5 - 3.5 {\rm \,kpc}$, then it evens out somewhat, until $r \approx 5{\rm \, kpc}$, where the decrease becomes more prominent again.

In the case of each of the three galaxies, the shapes of CO(2-1) and $\Sigma_\text{mol}$ mostly follow \CII\ throughout the profile. This similarity breaks down for NGC\,3627, owing to a significant central \CII\ deficit for this galaxy as well as a somewhat milder peak of CO(2-1) at the edges of its bar. The \CII\ drop is somewhat steeper throughout the profile in all three cases. However, for NGC\,3627 and NGC\,6946, it picks up slightly again at the very edges, unlike for NGC\,4321, which may be attributed to the fact that the observational area for NGC\,4321 does not encompass the farthest parts of this galaxy. In all three cases, \HI, and thus $\Sigma_\mathrm{atom}$, assume a completely different shape and do not seem to be affected by \CII, a feature prominent in the intensity maps as well.

\section{The performance of \CII\ as a resolved diagnostic of SFR and gas mass}\label{sec:analysis}

\subsection{\CII\ as an SFR tracer}

 \begin{table*}[h]
     \centering
\caption{Fitting slope and intercept values for the $\Sigma_\text{\CII} - \Sigma_\text{SFR}$ relation (Eq.\,\ref{eq:sfrcii}).} 

  \begin{tabular}{l c c c c c c }
      \hline
    \hline
    \multirow{2}{*}{Galaxy} &
    \multirow{2}{*}{Environment} &
    \multicolumn{2}{c}{Uncorrected} &
    \multicolumn{2}{c}{Corrected}

 \\
     & &    Slope (a) & Intercept (b)&    Slope (a) & Intercept (b) \\
    \hline
      All & All& $ 1.40 \pm 0.03$&$ -57.0 \pm0.2 $ & $1.27 \pm 0.03$ & $-52.0 \pm 0.1$ \\
    \hline
      NGC\,3627 & All& $1.68 \pm 0.06$ & $-68.8 \pm 0.2$  & $1.49 \pm 0.04$ & $-61.4 \pm 0.2$  \\
      & Centre& $1.9 \pm 0.5$& $-78 \pm 2$& $1.3 \pm 0.2$ & $-54.5  \pm 0.9$  \\
      & Outer disc & $1.57 \pm 0.05$ &$-64.4 \pm 0.2$& $1.44 \pm 0.04$  & $-59.4 \pm 0.2$ \\
      & Arm & $1.28 \pm 0.08$& $-53.0 \pm 0.4$ & $1.15 \pm 0.06$ & $-47.5 \pm 0.4$ \\
      & Interarm & $ 1.67 \pm 0.07$& $-68.6 \pm 0.3$ & $1.54 \pm 0.06$ & $-63.1 \pm 0.3$  \\
          \hline
      NGC\,4321 & All & $1.01 \pm 0.14$& $-41.5 \pm 0.9$& $0.97 \pm 0.09$ & $-39.6 \pm 0.6$  \\
      & Centre & $1.3 \pm 0.1$ &$-54.8 \pm 0.7$ & $1.2 \pm 0.1$ & $-47.6 \pm 0.6$\\
      & Outer disc &$ 0.36 \pm 0.05$&$-15.8 \pm 0.9$ & $0.36 \pm 0.05$  & $-15.7 \pm 0.9 $\\
      & Arm &$0.9 \pm 0.2$ &$-38.5 \pm 0.6$ &$1.0 \pm 0.2$ &$ -39.8 \pm 0.5 $ \\
      & Interarm & $0.23 \pm 0.04$& $-11 \pm 1$& $0.23 \pm 0.04$ & $-11 \pm 1$ \\
    \hline
      NGC\,6946 & All &$1.43 \pm 0.02$&$-58.4\pm 0.1$ & $1.27 \pm 0.02$ & $-51.8 \pm 0.1 $ \\
      & Centre &$2.2\pm 0.1$&$-87.8 \pm 0.4$ & $1.56 \pm 0.07$ & $-63.7 \pm 0.3 $ \\
      & Outer disc & $1.34 \pm 0.04$&$-54.9 \pm 0.2$ &$1.24 \pm 0.03$ & $-50.8 \pm 0.2 $\\
      & Arm &$1.29 \pm 0.05$&$-52.8 \pm 0.4$ & $1.15 \pm 0.03$ & $-47.2 \pm 0.3 $\\
      &  Interarm &$1.7 \pm 0.1$& $-69.3 \pm 0.6$& $1.7 \pm 0.1$ & $-67.5 \pm 0.5$ \\

      \hline \hline
  \end{tabular}
    \label{tab:all_coeff}
\end{table*}

\begin{figure*}
    \centering
    \includegraphics[width=0.8\textwidth]{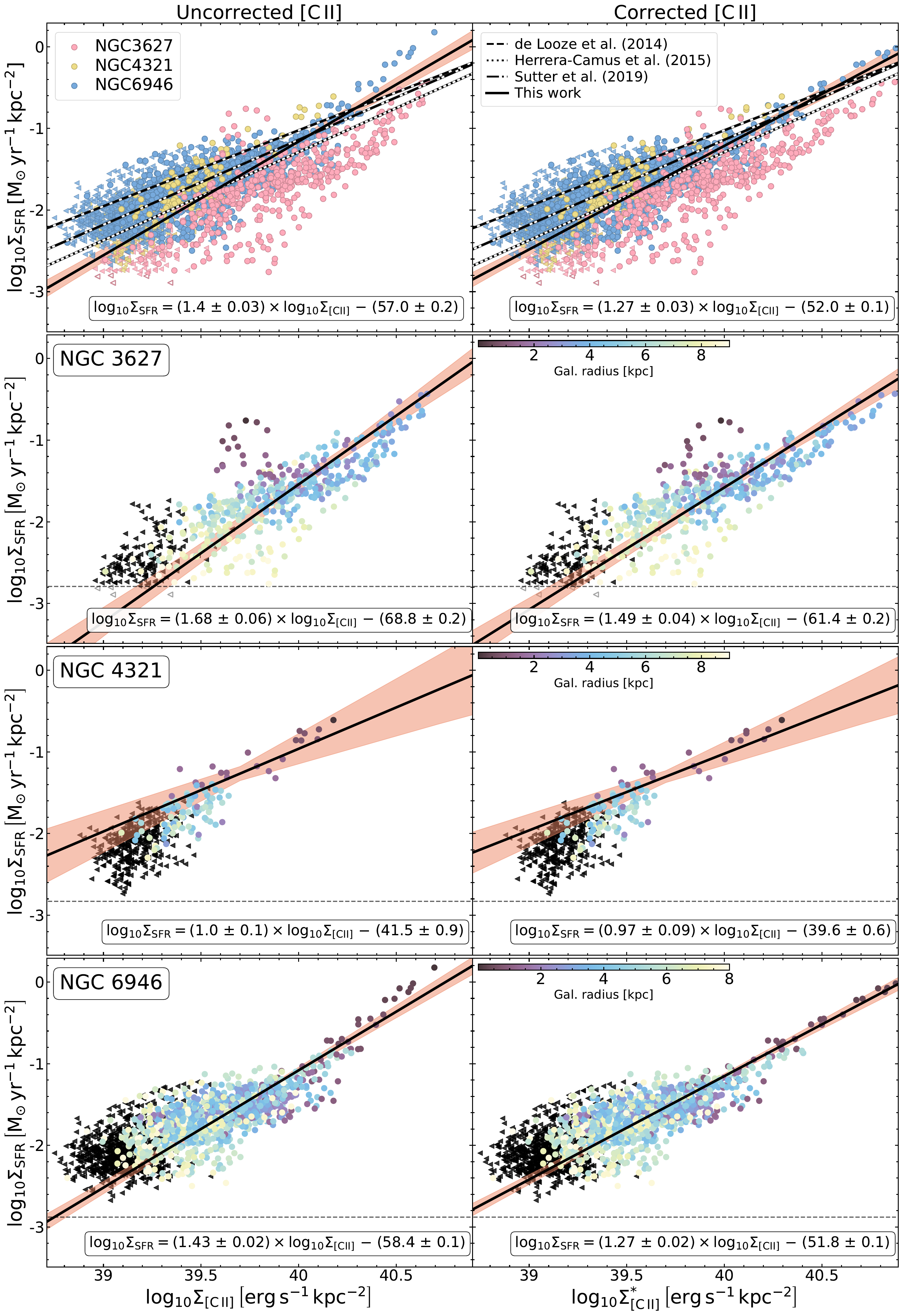}
    \caption{$\Sigma_\text{SFR}$ as a function of $\Sigma_\text{\CII}$ (left) or as IR colour-adjusted $\Sigma^*_\text{\CII}$, as described by \citet{Herrera} (right). The topmost row shows the fit for the entire dataset, followed by fits for each single galaxy: NGC\,3627, NGC\,4321, and finally NGC\,6946. Each point covers the half beam distance 0.50\,kpc for NGC\,3627, 0.67\,kpc for NGC\,4321, and 0.32\,kpc for NGC\,6946. Triangles represent points of upper limits, while circles represent actual measurements for points where \CII\ S/N $\geq$ 3. Empty points below the dashed horizontal line represent values below $3\, \times {\rm\,RMS}$\ noise for $\Sigma_\text{SFR}$ and are not included in the regression analysis. The fit is shown with a solid black line, and the red filled area around it marks the 3-$\sigma$ error. Dashed, dash dotted, and dotted lines (see legend) show fits from \citet{Herrera}, \citet{Sutter} (\textit{Herschel} KINGFISH and Beyond the Peak), and \citet{de_looze_2014} (\textit{Herschel} Dwarf Galaxy Survey).}
    \label{fig:mcmc_all_one}
\end{figure*}

 In this section we test the potential of \CII\ to trace the SFR throughout different environments within galaxies and from galaxy to galaxy. In Fig.\,\ref{fig:mcmc_all_one} we plot the $\Sigma_\text{SFR}$ as a function of \CII\ surface density ($\Sigma_\text{\CII}$) for all of the regions within the three galaxies. All environments of all galaxies taken together spread over the two orders of magnitude in $\Sigma_\text{\CII}$ and $\Sigma_\text{SFR}$. Values where the \CII\ ${\rm S/N} < 3$ were calculated as three times the uncertainty upper limits, marked as triangles, and included in the analysis. The values of $\Sigma_\text{SFR}$ that are less than three times the RMS noise are plotted with open circles and not included in the analysis.

Using the \texttt{Linmix}\footnote{https://linmix.readthedocs.io/en/latest/} package \citep{Kelly_2007}, we performed the Monte Carlo linear regression analysis for all of the galaxies together. We observed the expected behaviour of increasing $\Sigma_\text{SFR}$ for increasing $\Sigma_\text{\CII}$ found by several studies in the literature, such as \citet{de_looze_2014}, \citet{Herrera}, and \citet{Sutter}, which are also shown in Fig.\,\ref{fig:mcmc_all_one}. Our fit for the $\Sigma_\text{\CII} -  \Sigma_\text{SFR}$ relationship (solid black line in Fig.\,\ref{fig:mcmc_all_one}, top), which we derived here for all of the environments over all galaxies, taken together, is steeper than those in the literature.

For example, \citet{De_Looze_2011} first calibrated the SFR using \CII\ with a sample of galaxies classified as \HII\ regions, starbursts, or low-ionisation nuclear emission-line region galaxies (LINERs) and then expanded to a study of 48 dwarf galaxies \citep{de_looze_2014} from the \textit{Herschel} Dwarf Galaxy Survey \citep{madden_dgs} covering a wide range of metallicities -- from $12+\log({\rm O/H}) = 8.43$ to 7.14 -- and distances (several kiloparsecs to 191 Mpc). Using GALEX FUV and WISE 24\,\mic\ as a star formation tracer, the authors calibrated the SFR in \citet{De_Looze_2011} using \CII\ as
\begin{equation}
\begin{split}
   \log_{10} \Sigma_\text{SFR} &\left[\rm M_{\odot} \, yr^{-1}\, pc^{-2}\right]= \left(- 6.99 \pm 0.35\right) + \\ & \left(0.93 \pm 0.06\right) \times \log_{10}  \Sigma_\text{\CII} \left[{\rm L_{\odot}\, kpc^{-2}}\right] \quad .
    \end{split}
\end{equation}

In another study, \citet{Sutter} used 60 spatially resolved ($180-1700{\rm \,pc}$ and $200-2100 {\rm \,pc}$ for \CII\ 158\,\mic\ and [NII] 205\,\mic, respectively) normal star-forming Local Universe Galaxies ($3-30 {\rm \,Mpc}$) from the KINGFISH survey \citep{Kennicutt_2011} covering a wide range of metallicities. The authors noted the presence of the \CII\ deficit as being most prominent in the ionised phases of the ISM in their galaxy sample. Their relationship between the \CII\ line and SFR as traced by GALEX FUV and WISE 24\,\mic\ for combined ionised and neutral \CII\ emission for all individual regions is described as
\begin{equation}
\begin{split}
  \log_{10} \Sigma_\text{SFR} & \left[{\rm M_{\odot} \, yr^{-1}\, kpc^{-2}}\right] = -42.74 +  \\ & \left(1.04\pm 0.053 \right)  \times \log_{10} \Sigma_\text{\CII} \left[{\rm erg\,s^{-1}\, kpc^{-2}}\right] \quad.
   \end{split}
\end{equation}

One method to derive an accurate \CII-SFR calibration has been proposed by \citet{Herrera}. They used a \textit{Herschel} KINGFISH sample of 46 nearby galaxies with luminosities spanning $L_\text{TIR} \approx 10^{7.6} - 10^{11} {\rm \,L_{\odot} }$ and distances from $2.8$ to $26.5 {\rm \,Mpc}$, covering a large range of spatial scales from $0.2 - 1.5 {\rm \,kpc}$ (with the median value of $0.6 \pm 0.3 {\rm \,kpc}$). In order to increase the robustness of their \CII-SFR relation, they adjusted their SFR-\CII\ relation based on infrared colours, which they applied to their KINGFISH galaxy sample, and gave a relationship between adjusted \CII\ and SFR as
\begin{equation}
    \begin{split}
      \log_{10} \Sigma_\text{SFR} & \left[{\rm M_{\odot}\, yr^{-1}\, kpc^{-2}}\right] = - 1.29 +  \\ & 1.08 \times \left( \log_{10} \Sigma_\text{\CII} \left[{\rm erg\,s^{-1}\, kpc^{-2}}\right]-40\right)\quad .
    \end{split}
\end{equation}

We repeated the same procedure for each galaxy individually, colour-coded by radius, as well as on different environments within each galaxy, fitting the relationship in the following cases: 
\begin{itemize}
    \item the entire galaxy,
    \item the centre (representing both the galactic centre and bar),
    \item the arm,
    \item the interarm, and 
    \item the outer region (representing the arm and interarm region combined),
\end{itemize} 
in the form of
\begin{equation}
\begin{split}
    \log_{10}\ \Sigma_\text{SFR} &{\rm \,[M_{\odot} \,yr^{-1} \,kpc^{-2}]} = b {\rm \ } +
    \\ & a \times \log_{10}\ \Sigma_\text{\CII}{\rm\,[erg \,s^{-1}\, kpc^{-2}}] \quad .
\end{split}
    \label{eq:sfrcii}
\end{equation}
Slopes and intercept values that satisfied Eq.\,(\ref{eq:sfrcii}) for all our fits, with and without the IR-colour correction for \CII\ as per \citet{Herrera}, are listed in Table \ref{tab:all_coeff}.

When accounting for the whole galaxy, NGC\,4321 shows the shallowest relationship between $\Sigma_\text{SFR}$ and $\Sigma_\text{\CII}$, and NGC\,3627 shows the steepest. The highest $\Sigma_\text{SFR}$ and $\Sigma_\text{\CII}$ in the inner parts of the galaxy belong to NGC\,6946, followed by NGC\,4321, pointing to these two galaxies as being a better representation of the galaxy sample explored earlier in the literature. The \citet{Herrera} prescription has corrected the \CII\ deficit in them well.
 
On the other hand, the innermost points of the central region of NGC\,3627 show a very strong \CII\ deficit, which is not corrected by the \citet{Herrera} method applied here. The area brightest in the \CII\ belongs to the ends of the bars, which also show a high SFR, as seems to be the case with barred spiral galaxies \citep{Considere_2000}. The more distant regions lie on the lower end of the distribution for both SFR and \CII. This feature is similar in all three galaxies, and can be easily explained with the fact that the outermost regions mostly sampled the diffuse interarm environment in each galaxy. Several interarm points closer to the galactic centre are bright in \CII\ and have a high SFR.
 
In NGC\,3627 and NGC\,6946, the arm region shows the lowest slope of the fit (Table \ref{tab:all_coeff}). There, the SFR does not seem to contribute to the \CII\ emission as highly as, for instance, the central and interarm regions that follow the fit made across the full galaxy.

The outer environment within NGC\,4321 appears to differ from the two other galaxies in our sample. NGC\,4321 is also the most distant galaxy; thus our individually sampled regions cover larger linear areas. Being more distant, the \CII\ is fainter and therefore has a low number of \CII\ detections where S/N $\geq $ 3, meaning the SFR-\CII\ relation is less constrained in this environment.

Extending the analyses of the only other fully mapped individual galaxies, NGC\,6946 by \citet{Bigiel_2020} and NGC\,7331 by \citet{Sutter_2022}, this study compares in   detail the resolved SFR-\CII\ relation across the entire extent of the optical discs within three nearby galaxies, comparing variations towards different galaxy regions, as opposed to the local patches and/or radial strips typically covered with \textit{Herschel} \citep[e.g.][]{Cormier_2010,Kennicutt_2011,Herrera,de_Blok,Lapham_2017,Sutter}. The scatter in the SFR-\CII\ relation indicates that the use of \CII\ as an SFR tracer becomes less straightforward on resolved sub-kiloparsec ($\sim 0.5-1 {\rm \,kpc}$) scales, with a typical uncertainty of $0.75{\rm\,dex}$ compared to the global recipes from \citet{de_looze_2014} ($0.42{\rm\,dex}$) and \citet{Herrera} ($0.21{\rm\,dex}$). The spread seems mostly driven by galaxy-to-galaxy variations rather than local variations; that is, the spread within one galaxy is lower compared to the spread of the three galaxies combined. This result is similar to that inferred for the resolved Dwarf Galaxy Survey galaxies \citep{de_looze_2014}, and it was partly, but not entirely, alleviated by combining \CII\ with the emission of other important gas cooling lines (e.g. $\left[\ion{O}{i}\right]\ 63{\rm \,\mu m}$, $\left[\ion{O}{iii}\right]\ 88{\rm \, \mu m}$).

NGC\,3627 exhibits some observational characteristics that distinguish it from NGC\,4321 and NGC\,6936, most notably the absence of bright \CII\ emission from the nuclear region, where significant CO and TIR concentrations reside.  While the central region shows the lowest values of \CII/TIR and \CII/CO within the galaxy, these ratios increase by almost an order of magnitude towards the gas-rich star-forming bars (Fig.\,\ref{fig:all_three_binning_combined}). This likely indicates the influence of the central source dominated by an AGN in this galaxy. The relative deficit of \CII\ emission is characteristic of AGNs, which also contribute to the luminous \ltir. Consequently, application of \CII\ as an SFR tracer where the AGN may have an important influence, is not expected to be effective \citep[e.g.][]{Herrera_2018,Sutter}. This effect is not apparent in NGC\,4321, which also hosts an AGN. More detailed studies on sub-kiloparsec scales across full disc galaxies are necessary to determine the viability of using \CII\ as an SFR tracer within AGN-host galaxies beyond the influence of a central AGN.

While significant, our sample consists of only three galaxies, was derived from full disc observations instead global values, and shows prominent differences in the fits between various environments within galaxies themselves. Understanding the reason for these differences requires further study.
In the remainder of this section, we investigate several scenarios that could lead to an increased scatter in the resolved SFR-\CII\ relation.

\subsection{\CII\ as a tracer of molecular gas}\label{sec:mol_gas_tracer} 
\begin{figure}
    \centering
    \includegraphics[width=\columnwidth]{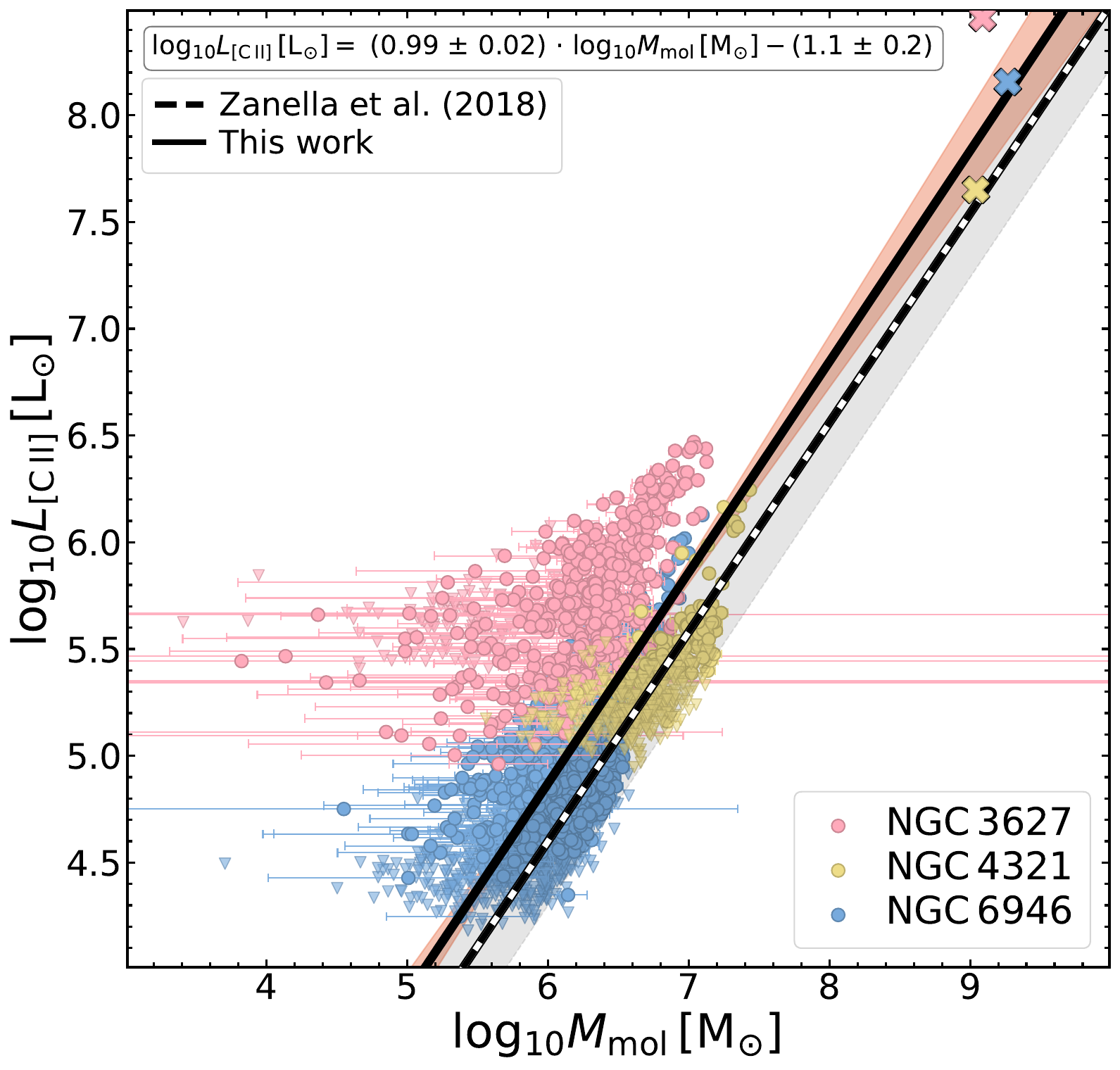}
    \caption{\CII\ luminosity as a function of $M_\text{mol}$ for galaxies NGC\,3627, NGC\,4321, and NGC\,6946. Each point covers  the half beam distance (0.50\,kpc for NGC\,3627, 0.67\,kpc for NGC\,4321, and 0.32\,kpc for NGC\,6946). Triangles represent points of upper limits, where \CII\ ${\rm S/N} < 3$, while circles represent actual measurements for points where \CII\ ${\rm S/N} \geq 3$, with error bars representing $\pm 1\sigma$ uncertainty. The solid black line shows the fit from our dataset, and the red filled area marks the 3-$\sigma$ error, while the dashed line marks the fit from Eq.\,(\ref{eq:zanella}) \citep{Zanella_2018}, along with their standard deviation shaded in grey. Large crosses in the upper-right corner of the plot represent integrated values for $L_\text{\CII}$ and $M_\text{mol}$, where \CII\ ${\rm S/N} \geq 3$, colour coded by galaxy.}
    \label{fig:zanella}
\end{figure}

\begin{table}[ht]
        \caption{Fitting slope and intercept values for the $M_\text{mol} - L_\text{\CII}$ relation (eq.\,\ref{eq:zanela_my}).}
\centering
  \begin{tabular}{l c c c }
      \hline
    \hline
           \multirow{2}{*}{Galaxy} &
    \multirow{2}{*}{Environment} &
    \multirow{2}{*}{Slope (a)} &
    \multirow{2}{*}{Intercept (b)}\\ \\

        \hline

       All & All & $  0.99 \pm 0.02  $ & $ -1.1 \pm 0.2  $\\
       \hline
       NGC\,3627 & All &$ 0.73 \pm 0.04  $ &$ 1.2 \pm 0.3 $ \\
        & Centre & $ 1.0 \pm 0.2  $&$ -1 \pm 2 $ \\
        & Outer disc & $ 0.71 \pm 0.04  $ &$ 1.3 \pm 0.3  $ \\
        & Arm & $  0.9 \pm 0.1  $&$ 0.3 \pm 1.0 $ \\
        & Interarm &$ 0.70 \pm 0.07  $ & $ 1.4\pm 0.4 $\\
\hline
NGC\,4321 & All & $ 1.9 \pm 0.2 $& $ -8 \pm 1 $ \\

        & Centre & $ 1.3 \pm 0.2  $&$ -4 \pm 1 $ \\
        & Outer disc & $ 2.3\pm 0.4  $ &$ -11 \pm 3  $ \\
        & Arm & $ 1.2 \pm 0.2  $&$  -3 \pm 2 $ \\
        & Interarm &$ 3.8 \pm 0.8  $ &$ -21\pm 5  $ \\
 \hline
 NGC\,6946 & All & $ 0.83 \pm 0.02  $&$ 0.0\pm 0.2  $ \\

        & Centre &$  0.76\pm 0.06 $ &$ 0.6 \pm 0.4  $ \\
        & Outer disc & $ 0.68 \pm 0.04 $&$ 0.9 \pm 0.2  $ \\
        & Arm & $  0.69 \pm 0.04 $&$ 0.8 \pm 0.2 $ \\
        & Interarm & $ 0.48 \pm 0.04  $ & $ 2.0 \pm 0.3  $\\   
        \hline \hline
    \end{tabular}

    \label{tab:ciimol}
\end{table}

A number of earlier studies have shown that $\approx 60-85 \%$ of \CII\ total luminosity originates from molecular clouds \citep{Accurso_2016,Olsen_2017}. \citet{Zanella_2018} used a sample of ten unresolved main-sequence galaxies at $z\approx2$ with stellar masses $\approx 10^{10}-10^{10.9} {\rm \,M_{\odot}}$ and SFRs $\approx 35-115 {\rm \,M_{\odot} \,yr^{-1}}$. They find a statistically significant (Spearman coefficient $\rho = 0.97$, $0.3 {\rm \,dex}$ scatter) linear $L_\text{\CII}-M_\text{mol}$ relationship:
\begin{equation}
\begin{split}
    \log_{10} L_\text{\CII} \left[{\rm L_{\odot}}\right] = & \left(0.98 \pm 0.02\right)\times \log_{10} M_\text{mol}\left[{\rm M_{\odot}}\right] \\&- \left(1.28\pm 0.21\right)\quad.
    \end{split}
\label{eq:zanella}
\end{equation}

In Fig.\,\ref{fig:zanella}, we plot this relation against the spatially resolved full disc observations of our galaxy sample and obtain a fit of the entire dataset that satisfies the equation
\begin{equation}
\begin{split}
    \log_{10} L_\text{\CII} \left[{\rm L_{\odot}}\right] = &  a \times \log_{10} M_\text{mol}\left[{\rm M_{\odot}}\right] + b\quad.
    \end{split}
\label{eq:zanela_my}
\end{equation}

The first point to notice is the very wide spread of the $L_\text{\CII}-M_\text{mol}$ relationship throughout all of our galaxies and from galaxy to galaxy. For the most part, only the central \CII\ bright points from our galaxies tend to correspond with the relationship from \citet{Zanella_2018}. The remaining points from the full disc observations for all three galaxies contribute to a high scatter ($\sim 0.7{\rm \, dex}$). NGC\,3627 stands out with very large deviations from Eq.\,(\ref{eq:zanella}). While our study uses spatially varying $\alpha_\text{CO}$ factors, the \citet{Zanella_2018} study incorporates a constant Galactic $\alpha_\text{CO}$ value. The applied $\alpha_\text{CO}$ explains some of the shift of the points within the galaxies, as demonstrated in Appendix \ref{appendixb}, but not in all of the regions. This points to a more complex relationship than has been suggested from earlier, more limited, and often unresolved observations. Local conditions and the choice of $\alpha_\text{CO}$ play an important role in linking the \CII\ to the molecular gas.

We repeated the fitting procedure for each individual galaxy as well as for different environments within the galaxies, and we list all of the slopes and intercepts in Table \ref{tab:ciimol}. We also calculated the integrated $L_\text{\CII}$ and $M_\text{mol}$ for each galaxy using only the $\text{\CII} \geq 3 \times {\rm S/N}$ observations (upper-right corner in Fig.\,\ref{fig:zanella}). The integrated values of NGC\,4321 and NGC\,6946 conform with the fit from \citet{Zanella_2018}, while NGC\,3627 deviates in this case as well.

\begin{figure*}
    \centering
    \includegraphics[width=\textwidth]{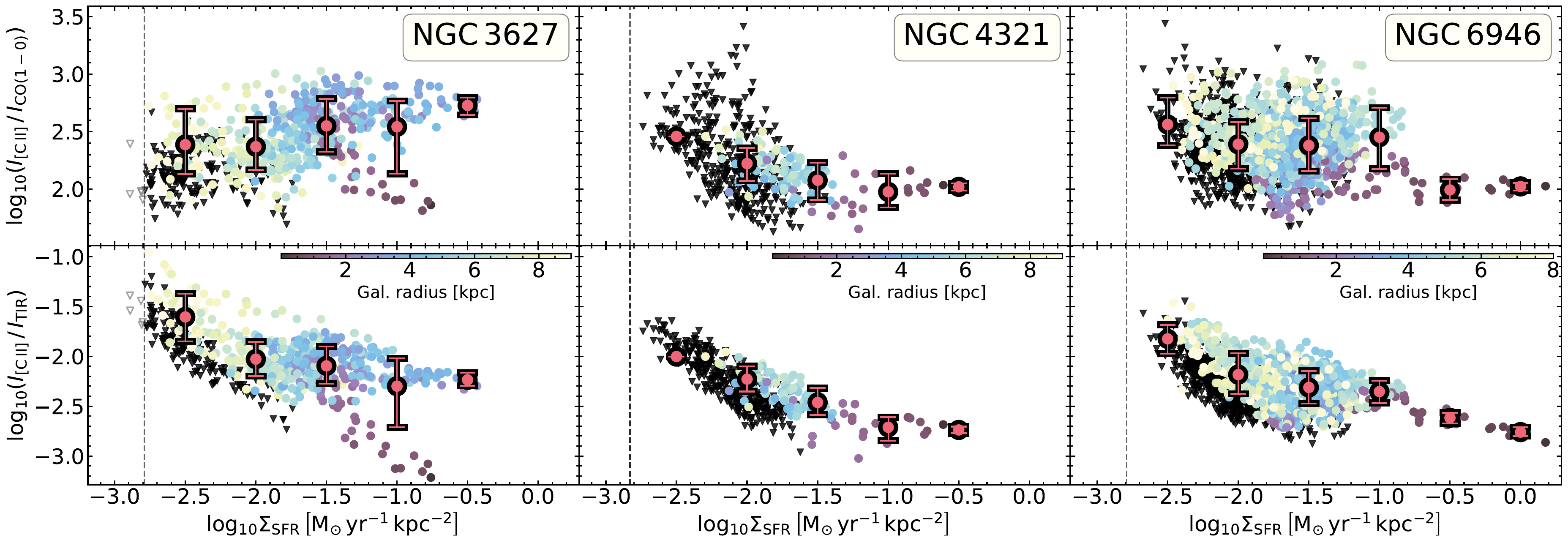}

    \caption{\CII/CO(1-0) and \CII/TIR line ratios as a function of $\Sigma_\text{SFR}$, colour-coded by radius. Grey triangles represent points of upper limits, where \CII\ ${\rm S/N} < 3$. Solid circles represent measurements for points where \CII\ ${\rm S/N} \geq 3$. All values left of the vertical line fall under the $3\, \times {\rm\,rms}$ for $\Sigma_\text{SFR}$. Large circles show means of bins with $\pm 1 \sigma$ error.
    }
    \label{fig:ciico_separate}
\end{figure*}

One of the issues that might complicate the use of \CII\ as an SFR tracer at resolved scales is its ambiguous origin. Carbon has an ionisation potential lower than hydrogen, which means that \CII\ is present in all but the coolest interstellar gas. Adding to the complexity, \CII\ also has a low critical density, and therefore arises from different density regimes. This makes it difficult to differentiate the contribution of \CII\ from various phases of the ISM in our observations. \citet{stacey_2010} proposed using a \CII/CO(1-0) ratio as a way to interpret the origin of the \CII\ fine-structure line since the CO molecule is only present in the molecular gas.

In Fig.\,\ref{fig:ciico_separate} we plot \CII/CO(1-0) intensity ratios as a function of $\Sigma_\text{SFR}$, colour-coded by galactocentric radius for each of the three galaxies. The large points with error bars mark the mean values of line ratios binned by $\Sigma_\text{SFR}$ with $\pm 1 \sigma$ error. In NGC\,3627, the mean remains relatively constant throughout the $\Sigma_\text{SFR}$ range. At the higher end, the central region forks out into the lower line ratio regime, as this galaxy is severely centrally \CII\ deficient.

NGC\,4321 and NGC\,6946 share a somewhat similar profile, with the central low \CII/CO(1-0) points falling to the high end of the $\Sigma_\text{SFR}$ regime. In all three cases, the low central line ratios as well as the high $\Sigma_\text{mol}$ values in the radial profile (Fig.\,\ref{fig:all_three_binning_combined}) point to \CII\ originating from the molecular gas, but for the rest of the galactic disc, there is too much scatter to draw definite conclusions. The only outliers are the atomic gas-rich outer bars of NGC\,3627 that show a high \CII/CO(1-0) ratio. Due to its past interaction with a neighbouring galaxy, this is a very active star-forming gas-rich area. It is possible that the high \CII/CO(1-0) line ratio for NGC\,3627 points to the atomic gas origins of \CII. However, in the case of all three galaxies, the scatter in the outer disc makes distinguishing the origin of \CII\ and its relationship to molecular gas and star formation extremely difficult.

\subsection{\CII\ as a tracer of photoelectric efficiency}

Apart from the uncertain origin of the \CII\ line, the \CII\ emission may be hampered by variations in the photoelectric efficiency. The smallest dust grains will predominantly contribute to the photoelectric heating of neutral interstellar gas, and variations in the photoelectric efficiency may lead to less gas heating for the same input level of UV photons from young stars, which will lead to less efficient collisional excitation of the \CII\ line and hence less \CII\ emission. The \CII/TIR ratio is often used as a tracer of the photoelectric efficiency, as it provides the ratio of the UV emission that went into heating the gas through the photoelectric effect (in case \CII\ is the dominant coolant in the neutral gas phase) and the energy that went into heating the dust.

We modelled $\Sigma_\text{SFR}$ in order to trace both the UV photons emitted by young bright stars directly via GALEX FUV 154\,nm as well as those already processed by the surrounding dust and re-emitted in the thermal IR ($10-300  {\rm \, \mu m}$) with WISE4 22\,\mic. Comparing the thus modelled $\Sigma_\mathrm{SFR}$ to \ltir\ emitted by the dust should give us better insight into the relationship between the \CII\ as an SFR tracer and the impact of a possible \CII\ deficit on the SFR-\CII\ relation. In Fig.\,\ref{fig:ciico_separate} we plot \CII/TIR ratio against $\Sigma_\text{SFR}$, and we observed that the \CII/TIR ratio covers a wide range ($\sim 10^{-3.8}-10^{-0.8}$ for NGC\,3627; however, that range falls to $\sim 10^{-3}-10^{-2}$ for NGC\,4321 and NGC\,6946), with the highest \CII/TIR ratios belonging to the mostly low star-forming outer regions of galactic discs. In \citet{Croxall_2012}, \CII/TIR for the galaxies NGC\,1097 and NGC\,4559 falls in the smaller range ($10^{-3}-10^{-2}$), showing the need for full disc \CII\ observations, as the earlier observations with \textit{Herschel} were focused on the central regions and strips of galaxies.
 
 Values of the ratio on the lower end of the SFR are higher, owing to the fact that \ltir\ and $\Sigma_\text{SFR}$ trace related quantities. All three galaxies have several central points on the high side of $\Sigma_\text{SFR}$ that extend beyond the bulk of the values. In this case, points with a high $\Sigma_\text{SFR}$ for NGC\,6946 show the smallest amount of scatter. NGC\,3627 has a fork at the higher end of $\Sigma_\text{SFR}$, with one side being due to the strong centrally uncorrected \CII\ deficit that results in a low \CII/TIR ratio and the other having some of the highest \CII/TIR that belong to the gas-rich outer bars of the galaxy. Central high $\Sigma_\text{SFR}$ points for NGC\,4321 and NGC\,6946 show line ratios that fall between those two extremes seen in NGC\,3627.

The trends shown in Fig.\,\ref{fig:ciico_separate} demonstrate that the \CII/TIR ratio varies over two orders of magnitude, with the high scatter observed outside of the central region of each galaxy (with the exception of the outer bars of NGC\,3627, most likely owing to its unique history). The trends also indicate that the photoelectric efficiency drops towards regions of high $\Sigma_\text{SFR}$, which are also the galaxy environments that deviate most significantly from the SFR-\CII\ relation (see Fig.\,\ref{fig:mcmc_all_one}). This drop corresponds to the \CII\ deficit towards warm IR colours and high TIR luminosities \citep{Croxall_2012,Smith_2017,Sutter_2022}.

\subsection{Second-order contributing factors}
\label{QPAH.sec}

\begin{figure*}
    \centering
        \includegraphics[width=\textwidth]{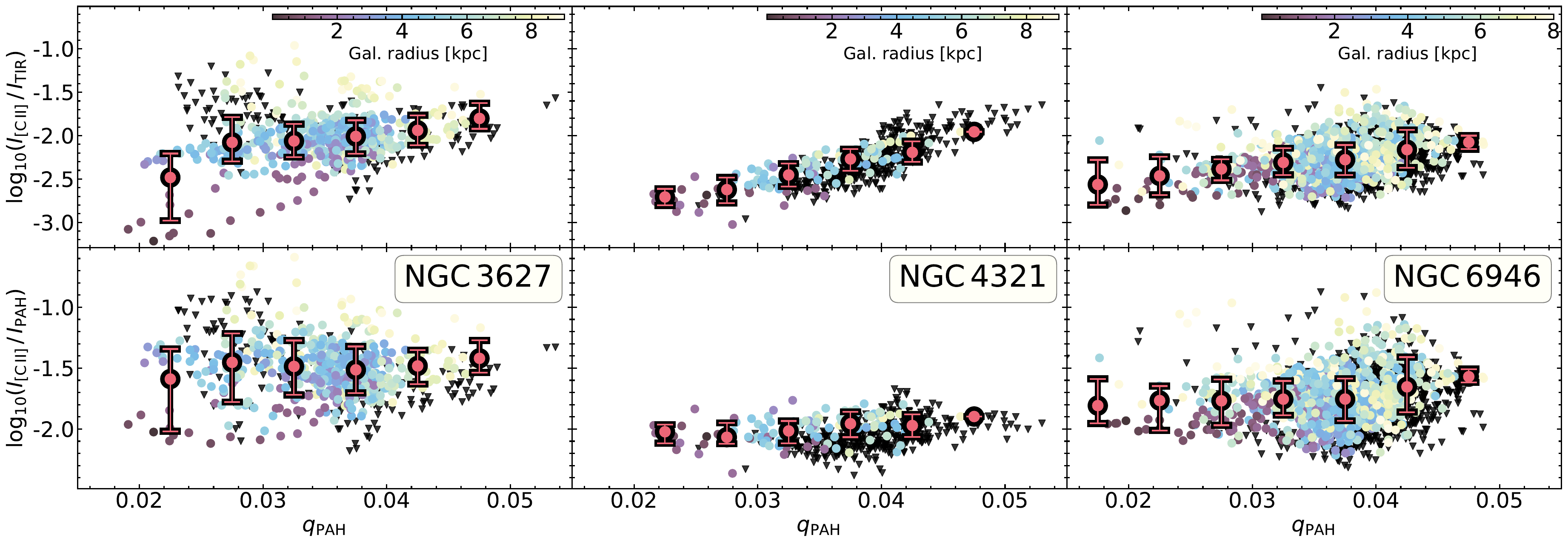}

    \caption{\CII/CO(1-0) and \CII/PAH line ratios as a function of $q_\text{PAH}$, colour-coded by radius. Grey triangles represent points of upper limits where \CII\ ${\rm S/N} < 3$. Solid circles represent measurements for points where \CII\ ${\rm S/N} \geq 3$. Large circles show means of bins with $\pm 1 \sigma$ error}
    \label{fig:gpah_fpdr}
\end{figure*}

Next, we investigate whether the drop in the \CII/TIR ratio can be linked to variations in the abundance of the smallest grains or variations in the contributions from old stars to the heating of dust accounting for a significant fraction of the TIR emission. As the smallest interstellar grains, consisting of less than $10^3$ carbon atoms \citep{munos_mateos}, PAHs are responsible for the bulk of photoelectric heating of interstellar gas \citep{helou_2001}. That makes the $q_\text{PAH}$ a good tracer of photoelectric efficiency. In Fig.\,\ref{fig:gpah_fpdr}, we show the distribution of the \CII/TIR ratio as a proxy for the photoelectric heating efficiency and the \CII/PAH ratio as a tracer of PAH emission, both as functions of $q_\text{PAH}$.

For all three galaxies, we find a drop in the \CII/TIR ratio in the central regions of the galaxies that seems to coincide with a lower $q_\text{PAH}$. This seems to suggest that the lower abundance of the smallest grains reduced the photoelectric heating efficiency and causes a drop in the \CII/TIR relation. It is of interest to note that this drop is seen in the central regions of all three galaxies regardless of whether they host an AGN (NGC\,3627 and NGC\,4321), whose strong radiation field would suppress photoelectric heating, or not (NGC\,6946). In an attempt to account for the reduced PAH fraction, we would expect that the \CII/TIR emission ratio exhibits a flatter behaviour since we then link the \CII\ emission directly to the smallest grains that contribute to the photoelectric effect. Indeed, this seems to bring the \CII/TIR ratios of the central regions in line with the lower range covered by the other environments in the disc.

\section{Summary and conclusion}\label{sec:summary}

In this work, we have presented the results of the SOFIA/FIFI-LS \CII\ full disc observations of galaxies NGC\,3627, NGC\,4321, and NGC\,6946. We combined them with ancillary observations of HERACLES CO(2-1), THINGS, and VIVA \HI\ lines as well as \textit{Herschel} and \textit{Spitzer} IR bands. We studied the radial trends of these lines and derived the properties of TIR intensity, SFR, atomic, and dark gas corrected molecular gas surface densities, and we obtained their line ratios, \CII/TIR, \CII/CO(2-1), as well as the ratios of $I_\text{\CII}/\Sigma_\text{mol}$ and $I_\text{\CII}/\Sigma_\text{atom}$.

We find that NGC\,3627 has a strong \CII\ deficit in the centre and a strong \CII\ intensity in the inner arms that then decreases throughout the disc. The \CII\ intensity for both NGC\,4321 and NGC\,6946 peaks at the centre, followed by a rapid decline that evens out somewhat for the remainder of the disc. The radial profiles of CO(2-1) intensity and TIR as well as the surface densities of molecular gas and SFR are somewhat similar to each other for each of the galaxies, with the greatest discrepancies being seen with the flatter CO(2-1) radial profile for NGC\,3627. Throughout the discs for NGC\,4321 and NGC\,6946, $\Sigma_\text{atom}$ is relatively constant, while the radial profile for NGC\,3627 is arch shaped.

The discrepancies in curve shapes are very prominent in radial profile ratios, where for NGC\,3627 \CII/TIR and $I_\text{\CII}/\Sigma_\text{mol}$ have a minimum in the centre and a maximum at the largest radius. However, \CII/CO(2-1) roughly follows the shape of the other two curves until $r \approx 4 {\rm \,kpc}$. For the other two galaxies, the three radial profile ratios are somewhat similar to each other.

Next, we calibrated the SFR against GALEX FUV and WISE4 bands and derived a relationship between $\Sigma_\text{\CII}$ and $\Sigma_\text{SFR}$ and compared it with the existing literature. We attempted to correct the so-called \CII\ deficit by applying the IR colour adjustment, which has a negligible effect on NGC\,3627 but works fairly well for the other two galaxies. The \CII-SFR relationship varies between the galaxies as well as between environments within each of the galaxies (from a slope of 0.29 for NGC\,4321 to 1.8 for NGC\,6946; both extremes belong to the interarm region). This deviation of the fit of the interarm region from the other environments within the galaxy is the most prominent in NGC\,4321, which can partly be explained with its distance since it affects the quality of the observations. The use of \CII\ as an SFR tracer within galaxies that may be hosting an AGN, such as NGC\,3627, may not be effective within the regions under the influence of the AGN.

Due to the ubiquity of \CII\ throughout multiple phases of the ISM, we tested whether it is possible to use \CII\ as a tracer of molecular gas. We modelled the molecular gas mass form the HERACLES CO(2-1) maps. We derived the relationship between $M_\text{mol}$ and \CII\ luminosity in our sample, and we compared it with the existing literature. We showed that, once again, the fit depends on the galaxy but also on the environment within the galaxy, as due to the scatter outside the central regions of NGC\,4321 and NGC\,6946, the fit comes with a high uncertainty ($\sim 0.70 {\rm \, dex}$). However, the global fits match the existing literature much better. Additionally, we attempted to interpret the origin of \CII\ via the \CII/CO(1-0) line ratio since the CO molecule is only present in molecular gas. We derived the CO(1-0) maps from the CO(2-1) intensity maps using line ratios calculated from HERACLES CO(2-1) and EMPIRE CO(1-0) maps on a lower resolution of 27\arcsec. Even with the corrected \CII\ deficit, the high star-forming central region shows a low \CII/CO(1-0) line ratio in our sample of galaxies. Generally, the highest \CII/CO(1-0) ratio comes from the low star-forming interarm gas. The inner arm region of NGC\,3627 is the only exception in that the high \CII/CO(1-0) ratio is correlated with regions of high star formation.

We also aimed at gaining insight into the efficiency of photoelectric heating of gas by PAH grains by comparing $\Sigma_\text{SFR}$ with \ltir. There is a low \CII/TIR ratio in the central region for all three galaxies. In the case of NGC\,3627, this is related to the very strong \CII\ deficit in the centre, but for the other two galaxies, the ratio is kept low by a high \ltir. Comparing these line ratios with models of $q_\text{PAH}$ explains some of the trends. The central regions of all three galaxies show a low $q_\text{PAH}$. NGC\,3627 and NGC\,4321 contain an AGN, which would lead to a low \CII/TIR ratio in the centre, as photoelectrically efficient PAHs are destroyed by radiation, so their heating does not have to be balanced by \CII\ cooling, thus keeping the \CII/TIR ratio low. Other important gas coolants that dominate in the denser regions (where the SFR and TIR are higher) are the $\left[\ion{O}{i}\right]$ 63 and 145\,\mic\ emission lines \citep[e.g.][]{Lebouteiller_2012}. 
However, the full disc observations of $\left[\ion{O}{i}\right]$ are not available for these three galaxies to test this concept.

In conclusion, our study utilising FIFI-LS on board SOFIA to map the far-IR $158{\rm \,\mu m}$ line from singly ionised carbon \CII\ in three nearby star-forming galaxies -- NGC\,3627, NGC\,4321, and NGC\,6946 -- has shed light on the intricate nature of \CII\ emission and its relationship with the processes within the ISM. 
While \CII\ observations are crucial for understanding the ISM in galaxies across various redshifts, our findings underscore the lack of a definitive constraint on the origins of \CII\ emission.
The observed variations in the relationship between the \CII\ fine structure line and SFR among the studied galaxies, as well as within different environments within each galaxy, emphasise the complexity of employing \CII\ as a tracer for star formation. 

Furthermore, our analysis reveals a nuanced connection between \CII\ and atomic and molecular gas tracers, making it challenging to pinpoint the specific environment giving rise to the \CII\ emission. 
To unravel this complexity, a larger sample size and additional observational tracers such as $\left[\ion{O}{i}\right]$ 63\,$\mu$m, an important coolant in warmer environments, or $\left[\ion{N}{ii}\right]$ 122 and 205\,$\mu$m, which are tracers of ionised medium, are required to estimate the contribution of \CII\ from the ionised phase of the ISM.

Contrary to previous suggestions within the extragalactic literature, which often focuses on small regions of galaxies or areas within the Milky Way or employs large apertures sampling diverse physical environments, our results highlight the intricate nature of using \CII\ as an SFR tracer or molecular gas mass tracer, at least at solar metallicities. In fact, the relationship between \CII\ and SFR or molecular gas can vary over at least 1\,dex or more, depending on local properties and what factor is used to convert from CO to molecular gas mass.

The complexity observed in the Milky Way's resolved observations is mirrored in our extragalactic study, emphasising the impact of local ISM conditions on interpreting \CII\ observations. 
In essence, our research underscores the need for a comprehensive approach involving a broader observational scope and additional tracers, especially with spatial resolution, in order to accurately disentangle the origins of \CII\ emission and its role in tracing star formation within galaxies.

\begin{acknowledgements}
We thank the referee for useful comments that helped to improve the quality of the manuscript.
Based on observations made with the NASA/DLR Stratospheric Observatory for Infrared Astronomy (SOFIA). SOFIA is jointly operated by the Universities Space Research Association, Inc. (USRA), under NASA contract NNA17BF53C, and the Deutsches SOFIA Institut (DSI) under DLR contract 50 OK 2002 to the University of Stuttgart. 
I.K. and M.B. gratefully acknowledge the financial support of the Flemish Fund for Scientific Research (FWO-Vlaanderen) through Research Projects G0G0420N and G0C4723N. 
I.D.L. gratefully acknowledges the support of the Research Foundation Flanders (FWO). 
N.S.S. gratefully acknowledges the support of the Research Foundation - Flanders (FWO Vlaanderen) grant 1290123N. 
S.v.d.G has received funding from the European Research Council (ERC) under the European Union’s Horizon 2020 research and innovation programme DustOrigin (ERC-2019-StG-851622), from the Bijzonder Onderzoeksfond (BOF) through the starting grant (BOF/STA/202002/006) and from the Flemish Fund for Scientific Research (FWO-Vlaanderen) through the research project G02382. 
A.N. gratefully acknowledges the support of the Belgian Federal Science Policy Office (BELSPO) for the provision of financial support in the framework of the PRODEX Programme of the European Space Agency (ESA) under contract number 4000143347.

\end{acknowledgements}

\bibliography{references}

\begin{thebibliography}{103}
\expandafter\ifx\csname natexlab\endcsname\relax\def\natexlab#1{#1}\fi

\bibitem[{{Accurso} {et~al.}(2017){Accurso}, {Saintonge}, {Bisbas}, \& {Viti}}]{Accurso_2016}
{Accurso}, G., {Saintonge}, A., {Bisbas}, T.~G., \& {Viti}, S. 2017, \mnras, 464, 3315

\bibitem[{{Anand} {et~al.}(2021{\natexlab{a}}){Anand}, {Lee}, {Van Dyk}, {Leroy}, {Rosolowsky}, {Schinnerer}, {Larson}, {Kourkchi}, {Kreckel}, {Scheuermann}, {Rizzi}, {Thilker}, {Tully}, {Bigiel}, {Blanc}, {Boquien}, {Chandar}, {Dale}, {Emsellem}, {Deger}, {Glover}, {Grasha}, {Groves}, {S. Klessen}, {Kruijssen}, {Querejeta}, {S{\'a}nchez-Bl{\'a}zquez}, {Schruba}, {Turner}, {Ubeda}, {Williams}, \& {Whitmore}}]{Anand2021b}
{Anand}, G.~S., {Lee}, J.~C., {Van Dyk}, S.~D., {et~al.} 2021{\natexlab{a}}, \mnras, 501, 3621

\bibitem[{{Anand} {et~al.}(2021{\natexlab{b}}){Anand}, {Rizzi}, {Tully}, {Shaya}, {Karachentsev}, {Makarov}, {Makarova}, {Wu}, {Dolphin}, \& {Kourkchi}}]{Anand2021a}
{Anand}, G.~S., {Rizzi}, L., {Tully}, R.~B., {et~al.} 2021{\natexlab{b}}, \aj, 162, 80

\bibitem[{{Aniano} {et~al.}(2020){Aniano}, {Draine}, {Hunt}, {Sandstrom}, {Calzetti}, {Kennicutt}, {Dale}, {Galametz}, {Gordon}, {Leroy}, {Smith}, {Roussel}, {Sauvage}, {Walter}, {Armus}, {Bolatto}, {Boquien}, {Crocker}, {De Looze}, {Donovan Meyer}, {Helou}, {Hinz}, {Johnson}, {Koda}, {Miller}, {Montiel}, {Murphy}, {Rela{\~n}o}, {Rix}, {Schinnerer}, {Skibba}, {Wolfire}, \& {Engelbracht}}]{Aniano_2020}
{Aniano}, G., {Draine}, B.~T., {Hunt}, L.~K., {et~al.} 2020, \apj, 889, 150

\bibitem[{{Bakes} \& {Tielens}(1994)}]{Bakes_1994}
{Bakes}, E.~L.~O. \& {Tielens}, A.~G.~G.~M. 1994, \apj, 427, 822

\bibitem[{{Bakes} \& {Tielens}(1998)}]{Bakes_1998}
{Bakes}, E.~L.~O. \& {Tielens}, A.~G.~G.~M. 1998, \apj, 499, 258

\bibitem[{{Bigiel} {et~al.}(2020){Bigiel}, {de Looze}, {Krabbe}, {Cormier}, {Barnes}, {Fischer}, {Bolatto}, {Bryant}, {Colditz}, {Geis}, {Herrera-Camus}, {Iserlohe}, {Klein}, {Leroy}, {Linz}, {Looney}, {Madden}, {Poglitsch}, {Stutzki}, \& {Vacca}}]{Bigiel_2020}
{Bigiel}, F., {de Looze}, I., {Krabbe}, A., {et~al.} 2020, \apj, 903, 30

\bibitem[{{Binggeli} {et~al.}(1985){Binggeli}, {Sandage}, \& {Tammann}}]{Binggeli}
{Binggeli}, B., {Sandage}, A., \& {Tammann}, G.~A. 1985, \aj, 90, 1681

\bibitem[{{Boselli} {et~al.}(2002){Boselli}, {Gavazzi}, {Lequeux}, \& {Pierini}}]{Boselli_2002}
{Boselli}, A., {Gavazzi}, G., {Lequeux}, J., \& {Pierini}, D. 2002, \aap, 385, 454

\bibitem[{{Chung} {et~al.}(2010){Chung}, {van Gorkom}, {Kenney}, {Crowl}, \& {Vollmer}}]{Chung_2010}
{Chung}, A., {van Gorkom}, J.~H., {Kenney}, J. D.~P., {Crowl}, H., \& {Vollmer}, B. 2010, \aj, 139, 2716

\bibitem[{{Colditz} {et~al.}(2018){Colditz}, {Beckmann}, {Bryant}, {Fischer}, {Fumi}, {Geis}, {Hamidouche}, {Henning}, {H{\"o}nle}, {Iserlohe}, {Klein}, {Krabbe}, {Looney}, {Poglitsch}, {Raab}, {Rebell}, {Rosenthal}, {Savage}, {Schweitzer}, \& {Vacca}}]{Colditz_2018}
{Colditz}, S., {Beckmann}, S., {Bryant}, A., {et~al.} 2018, Journal of Astronomical Instrumentation, 7, 1840004

\bibitem[{{Consid{\`e}re} {et~al.}(2000){Consid{\`e}re}, {Coziol}, {Contini}, \& {Davoust}}]{Considere_2000}
{Consid{\`e}re}, S., {Coziol}, R., {Contini}, T., \& {Davoust}, E. 2000, \aap, 356, 89

\bibitem[{{Cormier} {et~al.}(2010){Cormier}, {Madden}, {Hony}, {Contursi}, {Poglitsch}, {Galliano}, {Sturm}, {Doublier}, {Feuchtgruber}, {Galametz}, {Geis}, {de Jong}, {Okumura}, {Panuzzo}, \& {Sauvage}}]{Cormier_2010}
{Cormier}, D., {Madden}, S.~C., {Hony}, S., {et~al.} 2010, \aap, 518, L57

\bibitem[{{Crawford} {et~al.}(1985){Crawford}, {Genzel}, {Townes}, \& {Watson}}]{Crawford}
{Crawford}, M.~K., {Genzel}, R., {Townes}, C.~H., \& {Watson}, D.~M. 1985, \apj, 291, 755

\bibitem[{{Croxall} {et~al.}(2012){Croxall}, {Smith}, {Wolfire}, {Roussel}, {Sandstrom}, {Draine}, {Aniano}, {Dale}, {Armus}, {Beir{\~a}o}, {Helou}, {Bolatto}, {Appleton}, {Brandl}, {Calzetti}, {Crocker}, {Galametz}, {Groves}, {Hao}, {Hunt}, {Johnson}, {Kennicutt}, {Koda}, {Krause}, {Li}, {Meidt}, {Murphy}, {Rahman}, {Rix}, {Sauvage}, {Schinnerer}, {Walter}, \& {Wilson}}]{Croxall_2012}
{Croxall}, K.~V., {Smith}, J.~D., {Wolfire}, M.~G., {et~al.} 2012, \apj, 747, 81

\bibitem[{{de Blok} {et~al.}(2008){de Blok}, {Walter}, {Brinks}, {Trachternach}, {Oh}, \& {Kennicutt}}]{deBlok2008}
{de Blok}, W.~J.~G., {Walter}, F., {Brinks}, E., {et~al.} 2008, \aj, 136, 2648

\bibitem[{{de Blok} {et~al.}(2016){de Blok}, {Walter}, {Smith}, {Herrera-Camus}, {Bolatto}, {Requena-Torres}, {Crocker}, {Croxall}, {Kennicutt}, {Koda}, {Armus}, {Boquien}, {Dale}, {Kreckel}, \& {Meidt}}]{de_Blok}
{de Blok}, W.~J.~G., {Walter}, F., {Smith}, J. D.~T., {et~al.} 2016, \aj, 152, 51

\bibitem[{{De Looze} {et~al.}(2011){De Looze}, {Baes}, {Bendo}, {Cortese}, \& {Fritz}}]{De_Looze_2011}
{De Looze}, I., {Baes}, M., {Bendo}, G.~J., {Cortese}, L., \& {Fritz}, J. 2011, \mnras, 416, 2712

\bibitem[{{De Looze} {et~al.}(2014){De Looze}, {Cormier}, {Lebouteiller}, {Madden}, {Baes}, {Bendo}, {Boquien}, {Boselli}, {Clements}, {Cortese}, {Cooray}, {Galametz}, {Galliano}, {Graci{\'a}-Carpio}, {Isaak}, {Karczewski}, {Parkin}, {Pellegrini}, {R{\'e}my-Ruyer}, {Spinoglio}, {Smith}, \& {Sturm}}]{de_looze_2014}
{De Looze}, I., {Cormier}, D., {Lebouteiller}, V., {et~al.} 2014, \aap, 568, A62

\bibitem[{{den Brok} {et~al.}(2021){den Brok}, {Chatzigiannakis}, {Bigiel}, {Puschnig}, {Barnes}, {Leroy}, {Jim{\'e}nez-Donaire}, {Usero}, {Schinnerer}, {Rosolowsky}, {Faesi}, {Grasha}, {Hughes}, {Kruijssen}, {Liu}, {Neumann}, {Pety}, {Querejeta}, {Saito}, {Schruba}, \& {Stuber}}]{Den_Brok_2021}
{den Brok}, J.~S., {Chatzigiannakis}, D., {Bigiel}, F., {et~al.} 2021, \mnras, 504, 3221

\bibitem[{{D{\'\i}az-Santos} {et~al.}(2017){D{\'\i}az-Santos}, {Armus}, {Charmandaris}, {Lu}, {Stierwalt}, {Stacey}, {Malhotra}, {van der Werf}, {Howell}, {Privon}, {Mazzarella}, {Goldsmith}, {Murphy}, {Barcos-Mu{\~n}oz}, {Linden}, {Inami}, {Larson}, {Evans}, {Appleton}, {Iwasawa}, {Lord}, {Sanders}, \& {Surace}}]{Diaz_Santos_2017}
{D{\'\i}az-Santos}, T., {Armus}, L., {Charmandaris}, V., {et~al.} 2017, \apj, 846, 32

\bibitem[{{Draine} \& {Li}(2007)}]{DL07}
{Draine}, B.~T. \& {Li}, A. 2007, \apj, 657, 810

\bibitem[{{Efremov} \& {Moiseev}(2016)}]{Efremov_2016}
{Efremov}, Y.~N. \& {Moiseev}, A.~V. 2016, \mnras, 461, 2993

\bibitem[{{Erickson} \& {Davidson}(1993)}]{SOFIA}
{Erickson}, E.~F. \& {Davidson}, J.~A. 1993, Advances in Space Research, 13, 549

\bibitem[{{Fadda} {et~al.}(2023){Fadda}, {Colditz}, {Fischer}, {Vacca}, {Chu}, {Clarke}, {Klein}, {Krabbe}, {Minchin}, \& {Poglitsch}}]{cii_psf}
{Fadda}, D., {Colditz}, S., {Fischer}, C., {et~al.} 2023, \aj, 166, 237

\bibitem[{{Ferrara} {et~al.}(2019){Ferrara}, {Vallini}, {Pallottini}, {Gallerani}, {Carniani}, {Kohandel}, {Decataldo}, \& {Behrens}}]{Ferrara_2019}
{Ferrara}, A., {Vallini}, L., {Pallottini}, A., {et~al.} 2019, \mnras, 489, 1

\bibitem[{{Fischer} {et~al.}(2018{\natexlab{a}}){Fischer}, {Beckmann}, {Bryant}, {Colditz}, {Fumi}, {Geis}, {Hamidouche}, {Henning}, {H{\"o}nle}, {Iserlohe}, {Klein}, {Krabbe}, {Looney}, {Poglitsch}, {Raab}, {Rebell}, {Rosenthal}, {Savage}, {Schweitzer}, {Trinh}, \& {Vacca}}]{FIFI}
{Fischer}, C., {Beckmann}, S., {Bryant}, A., {et~al.} 2018{\natexlab{a}}, Journal of Astronomical Instrumentation, 7, 1840003

\bibitem[{{Fischer} {et~al.}(2018{\natexlab{b}}){Fischer}, {Beckmann}, {Bryant}, {Colditz}, {Fumi}, {Geis}, {Hamidouche}, {Henning}, {H{\"o}nle}, {Iserlohe}, {Klein}, {Krabbe}, {Looney}, {Poglitsch}, {Raab}, {Rebell}, {Rosenthal}, {Savage}, {Schweitzer}, {Trinh}, \& {Vacca}}]{Fischer_2018}
{Fischer}, C., {Beckmann}, S., {Bryant}, A., {et~al.} 2018{\natexlab{b}}, Journal of Astronomical Instrumentation, 7, 1840003

\bibitem[{{Galametz} {et~al.}(2013){Galametz}, {Kennicutt}, {Calzetti}, {Aniano}, {Draine}, {Boquien}, {Brandl}, {Croxall}, {Dale}, {Engelbracht}, {Gordon}, {Groves}, {Hao}, {Helou}, {Hinz}, {Hunt}, {Johnson}, {Li}, {Murphy}, {Roussel}, {Sandstrom}, {Skibba}, \& {Tabatabaei}}]{Galametz_2013}
{Galametz}, M., {Kennicutt}, R.~C., {Calzetti}, D., {et~al.} 2013, \mnras, 431, 1956

\bibitem[{{Garcia}(1993)}]{garcia_1993}
{Garcia}, A.~M. 1993, \aaps, 100, 47

\bibitem[{{Graci{\'a}-Carpio} {et~al.}(2011){Graci{\'a}-Carpio}, {Sturm}, {Hailey-Dunsheath}, {Fischer}, {Contursi}, {Poglitsch}, {Genzel}, {Gonz{\'a}lez-Alfonso}, {Sternberg}, {Verma}, {Christopher}, {Davies}, {Feuchtgruber}, {de Jong}, {Lutz}, \& {Tacconi}}]{Garcia_Carpio_2011}
{Graci{\'a}-Carpio}, J., {Sturm}, E., {Hailey-Dunsheath}, S., {et~al.} 2011, \apjl, 728, L7

\bibitem[{{Helou} {et~al.}(2001){Helou}, {Malhotra}, {Hollenbach}, {Dale}, \& {Contursi}}]{helou_2001}
{Helou}, G., {Malhotra}, S., {Hollenbach}, D.~J., {Dale}, D.~A., \& {Contursi}, A. 2001, \apjl, 548, L73

\bibitem[{{Herrera-Camus} {et~al.}(2015){Herrera-Camus}, {Bolatto}, {Wolfire}, {Smith}, {Croxall}, {Kennicutt}, {Calzetti}, {Helou}, {Walter}, {Leroy}, {Draine}, {Brandl}, {Armus}, {Sandstrom}, {Dale}, {Aniano}, {Meidt}, {Boquien}, {Hunt}, {Galametz}, {Tabatabaei}, {Murphy}, {Appleton}, {Roussel}, {Engelbracht}, \& {Beirao}}]{Herrera}
{Herrera-Camus}, R., {Bolatto}, A.~D., {Wolfire}, M.~G., {et~al.} 2015, \apj, 800, 1

\bibitem[{{Herrera-Camus} {et~al.}(2018){Herrera-Camus}, {Sturm}, {Graci{\'a}-Carpio}, {Lutz}, {Contursi}, {Veilleux}, {Fischer}, {Gonz{\'a}lez-Alfonso}, {Poglitsch}, {Tacconi}, {Genzel}, {Maiolino}, {Sternberg}, {Davies}, \& {Verma}}]{Herrera_2018}
{Herrera-Camus}, R., {Sturm}, E., {Graci{\'a}-Carpio}, J., {et~al.} 2018, \apj, 861, 95

\bibitem[{{Hollenbach} {et~al.}(1991){Hollenbach}, {Takahashi}, \& {Tielens}}]{Hollenbach_1991}
{Hollenbach}, D.~J., {Takahashi}, T., \& {Tielens}, A.~G.~G.~M. 1991, \apj, 377, 192

\bibitem[{{Jim{\'e}nez-Donaire} {et~al.}(2019){Jim{\'e}nez-Donaire}, {Bigiel}, {Leroy}, {Usero}, {Cormier}, {Puschnig}, {Gallagher}, {Kepley}, {Bolatto}, {Garc{\'\i}a-Burillo}, {Hughes}, {Kramer}, {Pety}, {Schinnerer}, {Schruba}, {Schuster}, \& {Walter}}]{Jim_nez_Donaire_2019}
{Jim{\'e}nez-Donaire}, M.~J., {Bigiel}, F., {Leroy}, A.~K., {et~al.} 2019, \apj, 880, 127

\bibitem[{{Kaufman} {et~al.}(1999){Kaufman}, {Wolfire}, {Hollenbach}, \& {Luhman}}]{Kaufman_1999}
{Kaufman}, M.~J., {Wolfire}, M.~G., {Hollenbach}, D.~J., \& {Luhman}, M.~L. 1999, \apj, 527, 795

\bibitem[{{Kelly}(2007)}]{Kelly_2007}
{Kelly}, B.~C. 2007, \apj, 665, 1489

\bibitem[{{Kennicutt} {et~al.}(2011){Kennicutt}, {Calzetti}, {Aniano}, {Appleton}, {Armus}, {Beir{\~a}o}, {Bolatto}, {Brandl}, {Crocker}, {Croxall}, {Dale}, {Donovan Meyer}, {Draine}, {Engelbracht}, {Galametz}, {Gordon}, {Groves}, {Hao}, {Helou}, {Hinz}, {Hunt}, {Johnson}, {Koda}, {Krause}, {Leroy}, {Li}, {Meidt}, {Montiel}, {Murphy}, {Rahman}, {Rix}, {Roussel}, {Sandstrom}, {Sauvage}, {Schinnerer}, {Skibba}, {Smith}, {Srinivasan}, {Vigroux}, {Walter}, {Wilson}, {Wolfire}, \& {Zibetti}}]{Kennicutt_2011}
{Kennicutt}, R.~C., {Calzetti}, D., {Aniano}, G., {et~al.} 2011, \pasp, 123, 1347

\bibitem[{{Koopmann} \& {Kenney}(2004)}]{Koopmann}
{Koopmann}, R.~A. \& {Kenney}, J. D.~P. 2004, \apj, 613, 866

\bibitem[{{Lagache} {et~al.}(2018){Lagache}, {Cousin}, \& {Chatzikos}}]{Lagache_2018}
{Lagache}, G., {Cousin}, M., \& {Chatzikos}, M. 2018, \aap, 609, A130

\bibitem[{{Lang} {et~al.}(2020){Lang}, {Meidt}, {Rosolowsky}, {Nofech}, {Schinnerer}, {Leroy}, {Emsellem}, {Pessa}, {Glover}, {Groves}, {Hughes}, {Kruijssen}, {Querejeta}, {Schruba}, {Bigiel}, {Blanc}, {Chevance}, {Colombo}, {Faesi}, {Henshaw}, {Herrera}, {Liu}, {Pety}, {Puschnig}, {Saito}, {Sun}, \& {Usero}}]{Lang2020}
{Lang}, P., {Meidt}, S.~E., {Rosolowsky}, E., {et~al.} 2020, \apj, 897, 122

\bibitem[{{Langer} \& {Pineda}(2015)}]{Langer_2015}
{Langer}, W.~D. \& {Pineda}, J.~L. 2015, \aap, 580, A5

\bibitem[{{Lapham} {et~al.}(2017){Lapham}, {Young}, \& {Crocker}}]{Lapham_2017}
{Lapham}, R.~C., {Young}, L.~M., \& {Crocker}, A. 2017, \apj, 840, 51

\bibitem[{{Le F{\`e}vre} {et~al.}(2020){Le F{\`e}vre}, {B{\'e}thermin}, {Faisst}, {Jones}, {Capak}, {Cassata}, {Silverman}, {Schaerer}, {Yan}, {Amorin}, {Bardelli}, {Boquien}, {Cimatti}, {Dessauges-Zavadsky}, {Giavalisco}, {Hathi}, {Fudamoto}, {Fujimoto}, {Ginolfi}, {Gruppioni}, {Hemmati}, {Ibar}, {Koekemoer}, {Khusanova}, {Lagache}, {Lemaux}, {Loiacono}, {Maiolino}, {Mancini}, {Narayanan}, {Morselli}, {M{\'e}ndez-Hern{\`a}ndez}, {Oesch}, {Pozzi}, {Romano}, {Riechers}, {Scoville}, {Talia}, {Tasca}, {Thomas}, {Toft}, {Vallini}, {Vergani}, {Walter}, {Zamorani}, \& {Zucca}}]{LeFevre_2020}
{Le F{\`e}vre}, O., {B{\'e}thermin}, M., {Faisst}, A., {et~al.} 2020, \aap, 643, A1

\bibitem[{{Lebouteiller} {et~al.}(2012){Lebouteiller}, {Cormier}, {Madden}, {Galliano}, {Indebetouw}, {Abel}, {Sauvage}, {Hony}, {Contursi}, {Poglitsch}, {R{\'e}my}, {Sturm}, \& {Wu}}]{Lebouteiller_2012}
{Lebouteiller}, V., {Cormier}, D., {Madden}, S.~C., {et~al.} 2012, \aap, 548, A91

\bibitem[{{Leech} {et~al.}(1999){Leech}, {V{\"o}lk}, {Heinrichsen}, {Hippelein}, {Metcalfe}, {Pierini}, {Popescu}, {Tuffs}, \& {Xu}}]{Leech_1999}
{Leech}, K.~J., {V{\"o}lk}, H.~J., {Heinrichsen}, I., {et~al.} 1999, \mnras, 310, 317

\bibitem[{{Leroy} {et~al.}(2019{\natexlab{a}}){Leroy}, {Sandstrom}, {Lang}, {Lewis}, {Salim}, {Behrens}, {Chastenet}, {Chiang}, {Gallagher}, {Kessler}, \& {Utomo}}]{Leroy2019}
{Leroy}, A.~K., {Sandstrom}, K.~M., {Lang}, D., {et~al.} 2019{\natexlab{a}}, \apjs, 244, 24

\bibitem[{{Leroy} {et~al.}(2019{\natexlab{b}}){Leroy}, {Sandstrom}, {Lang}, {Lewis}, {Salim}, {Behrens}, {Chastenet}, {Chiang}, {Gallagher}, {Kessler}, \& {Utomo}}]{Leroy_2019}
{Leroy}, A.~K., {Sandstrom}, K.~M., {Lang}, D., {et~al.} 2019{\natexlab{b}}, \apjs, 244, 24

\bibitem[{{Leroy} {et~al.}(2021{\natexlab{a}}){Leroy}, {Schinnerer}, {Hughes}, {Rosolowsky}, {Pety}, {Schruba}, {Usero}, {Blanc}, {Chevance}, {Emsellem}, {Faesi}, {Herrera}, {Liu}, {Meidt}, {Querejeta}, {Saito}, {Sandstrom}, {Sun}, {Williams}, {Anand}, {Barnes}, {Behrens}, {Belfiore}, {Benincasa}, {Be{\v{s}}li{\'c}}, {Bigiel}, {Bolatto}, {den Brok}, {Cao}, {Chandar}, {Chastenet}, {Chiang}, {Congiu}, {Dale}, {Deger}, {Eibensteiner}, {Egorov}, {Garc{\'\i}a-Rodr{\'\i}guez}, {Glover}, {Grasha}, {Henshaw}, {Ho}, {Kepley}, {Kim}, {Klessen}, {Kreckel}, {Koch}, {Kruijssen}, {Larson}, {Lee}, {Lopez}, {Machado}, {Mayker}, {McElroy}, {Murphy}, {Ostriker}, {Pan}, {Pessa}, {Puschnig}, {Razza}, {S{\'a}nchez-Bl{\'a}zquez}, {Santoro}, {Sardone}, {Scheuermann}, {Sliwa}, {Sormani}, {Stuber}, {Thilker}, {Turner}, {Utomo}, {Watkins}, \& {Whitmore}}]{Leroy2021a}
{Leroy}, A.~K., {Schinnerer}, E., {Hughes}, A., {et~al.} 2021{\natexlab{a}}, \apjs, 257, 43

\bibitem[{{Leroy} {et~al.}(2021{\natexlab{b}}){Leroy}, {Schinnerer}, {Hughes}, {Rosolowsky}, {Pety}, {Schruba}, {Usero}, {Blanc}, {Chevance}, {Emsellem}, {Faesi}, {Herrera}, {Liu}, {Meidt}, {Querejeta}, {Saito}, {Sandstrom}, {Sun}, {Williams}, {Anand}, {Barnes}, {Behrens}, {Belfiore}, {Benincasa}, {Be{\v{s}}li{\'c}}, {Bigiel}, {Bolatto}, {den Brok}, {Cao}, {Chandar}, {Chastenet}, {Chiang}, {Congiu}, {Dale}, {Deger}, {Eibensteiner}, {Egorov}, {Garc{\'\i}a-Rodr{\'\i}guez}, {Glover}, {Grasha}, {Henshaw}, {Ho}, {Kepley}, {Kim}, {Klessen}, {Kreckel}, {Koch}, {Kruijssen}, {Larson}, {Lee}, {Lopez}, {Machado}, {Mayker}, {McElroy}, {Murphy}, {Ostriker}, {Pan}, {Pessa}, {Puschnig}, {Razza}, {S{\'a}nchez-Bl{\'a}zquez}, {Santoro}, {Sardone}, {Scheuermann}, {Sliwa}, {Sormani}, {Stuber}, {Thilker}, {Turner}, {Utomo}, {Watkins}, \& {Whitmore}}]{Leroy_2021}
{Leroy}, A.~K., {Schinnerer}, E., {Hughes}, A., {et~al.} 2021{\natexlab{b}}, \apjs, 257, 43

\bibitem[{{Leroy} {et~al.}(2009){Leroy}, {Walter}, {Bigiel}, {Usero}, {Weiss}, {Brinks}, {de Blok}, {Kennicutt}, {Schuster}, {Kramer}, {Wiesemeyer}, \& {Roussel}}]{Leroy_2009}
{Leroy}, A.~K., {Walter}, F., {Bigiel}, F., {et~al.} 2009, \aj, 137, 4670

\bibitem[{{Leroy} {et~al.}(2013){Leroy}, {Walter}, {Sandstrom}, {Schruba}, {Munoz-Mateos}, {Bigiel}, {Bolatto}, {Brinks}, {de Blok}, {Meidt}, {Rix}, {Rosolowsky}, {Schinnerer}, {Schuster}, \& {Usero}}]{Leroy2013}
{Leroy}, A.~K., {Walter}, F., {Sandstrom}, K., {et~al.} 2013, \aj, 146, 19

\bibitem[{{Lord}(1992)}]{atran}
{Lord}, S.~D. 1992, {A new software tool for computing Earth's atmospheric transmission of near- and far-infrared radiation}, NASA Technical Memorandum 103957

\bibitem[{{Madden} {et~al.}(2020){Madden}, {Cormier}, {Hony}, {Lebouteiller}, {Abel}, {Galametz}, {De Looze}, {Chevance}, {Polles}, {Lee}, {Galliano}, {Lambert-Huyghe}, {Hu}, \& {Ramambason}}]{Madden_2020}
{Madden}, S.~C., {Cormier}, D., {Hony}, S., {et~al.} 2020, \aap, 643, A141

\bibitem[{{Madden} {et~al.}(2013){Madden}, {R{\'e}my-Ruyer}, {Galametz}, {Cormier}, {Lebouteiller}, {Galliano}, {Hony}, {Bendo}, {Smith}, {Pohlen}, {Roussel}, {Sauvage}, {Wu}, {Sturm}, {Poglitsch}, {Contursi}, {Doublier}, {Baes}, {Barlow}, {Boselli}, {Boquien}, {Carlson}, {Ciesla}, {Cooray}, {Cortese}, {de Looze}, {Irwin}, {Isaak}, {Kamenetzky}, {Karczewski}, {Lu}, {MacHattie}, {O'Halloran}, {Parkin}, {Rangwala}, {Schirm}, {Schulz}, {Spinoglio}, {Vaccari}, {Wilson}, \& {Wozniak}}]{madden_dgs}
{Madden}, S.~C., {R{\'e}my-Ruyer}, A., {Galametz}, M., {et~al.} 2013, \pasp, 125, 600

\bibitem[{{Maiolino} {et~al.}(2005){Maiolino}, {Cox}, {Caselli}, {Beelen}, {Bertoldi}, {Carilli}, {Kaufman}, {Menten}, {Nagao}, {Omont}, {Wei{\ss}}, {Walmsley}, \& {Walter}}]{Maiolino}
{Maiolino}, R., {Cox}, P., {Caselli}, P., {et~al.} 2005, \aap, 440, L51

\bibitem[{{Makarov} {et~al.}(2014){Makarov}, {Prugniel}, {Terekhova}, {Courtois}, \& {Vauglin}}]{Makarov2014}
{Makarov}, D., {Prugniel}, P., {Terekhova}, N., {Courtois}, H., \& {Vauglin}, I. 2014, \aap, 570, A13

\bibitem[{{Malhotra} {et~al.}(1997){Malhotra}, {Helou}, {Stacey}, {Hollenbach}, {Lord}, {Beichman}, {Dinerstein}, {Hunter}, {Lo}, {Lu}, {Rubin}, {Silbermann}, {Thronson}, \& {Werner}}]{Malhotra_1997}
{Malhotra}, S., {Helou}, G., {Stacey}, G., {et~al.} 1997, \apjl, 491, L27

\bibitem[{{Malhotra} {et~al.}(2001){Malhotra}, {Kaufman}, {Hollenbach}, {Helou}, {Rubin}, {Brauher}, {Dale}, {Lu}, {Lord}, {Stacey}, {Contursi}, {Hunter}, \& {Dinerstein}}]{Malhotra}
{Malhotra}, S., {Kaufman}, M.~J., {Hollenbach}, D., {et~al.} 2001, \apj, 561, 766

\bibitem[{{Martin} {et~al.}(2005){Martin}, {Fanson}, {Schiminovich}, {Morrissey}, {Friedman}, {Barlow}, {Conrow}, {Grange}, {Jelinsky}, {Milliard}, {Siegmund}, {Bianchi}, {Byun}, {Donas}, {Forster}, {Heckman}, {Lee}, {Madore}, {Malina}, {Neff}, {Rich}, {Small}, {Surber}, {Szalay}, {Welsh}, \& {Wyder}}]{galex}
{Martin}, D.~C., {Fanson}, J., {Schiminovich}, D., {et~al.} 2005, \apjl, 619, L1

\bibitem[{{Mathis} {et~al.}(1983){Mathis}, {Mezger}, \& {Panagia}}]{Mathis_1983}
{Mathis}, J.~S., {Mezger}, P.~G., \& {Panagia}, N. 1983, \aap, 128, 212

\bibitem[{{Moustakas} {et~al.}(2010){Moustakas}, {Kennicutt}, {Tremonti}, {Dale}, {Smith}, \& {Calzetti}}]{Moustakas_2010}
{Moustakas}, J., {Kennicutt}, Robert~C., J., {Tremonti}, C.~A., {et~al.} 2010, \apjs, 190, 233

\bibitem[{{Mu{\~n}oz-Mateos} {et~al.}(2009){Mu{\~n}oz-Mateos}, {Gil de Paz}, {Boissier}, {Zamorano}, {Dale}, {P{\'e}rez-Gonz{\'a}lez}, {Gallego}, {Madore}, {Bendo}, {Thornley}, {Draine}, {Boselli}, {Buat}, {Calzetti}, {Moustakas}, \& {Kennicutt}}]{munos_mateos}
{Mu{\~n}oz-Mateos}, J.~C., {Gil de Paz}, A., {Boissier}, S., {et~al.} 2009, \apj, 701, 1965

\bibitem[{{Nakagawa} {et~al.}(1998){Nakagawa}, {Yui}, {Doi}, {Okuda}, {Shibai}, {Mochizuki}, {Nishimura}, \& {Low}}]{Nakagawa_1998}
{Nakagawa}, T., {Yui}, Y.~Y., {Doi}, Y., {et~al.} 1998, \apjs, 115, 259

\bibitem[{{Olsen} {et~al.}(2017){Olsen}, {Greve}, {Narayanan}, {Thompson}, {Dav{\'e}}, {Niebla Rios}, \& {Stawinski}}]{Olsen_2017}
{Olsen}, K., {Greve}, T.~R., {Narayanan}, D., {et~al.} 2017, \apj, 846, 105

\bibitem[{{Peebles} \& {Nusser}(2010)}]{Peebles}
{Peebles}, P.~J.~E. \& {Nusser}, A. 2010, \nat, 465, 565

\bibitem[{{Petrosian} {et~al.}(1969){Petrosian}, {Bahcall}, \& {Salpeter}}]{Petrosian}
{Petrosian}, V., {Bahcall}, J.~N., \& {Salpeter}, E.~E. 1969, \apjl, 155, L57

\bibitem[{{Pierini} {et~al.}(1999){Pierini}, {Leech}, {Tuffs}, \& {Volk}}]{Pierini_1999}
{Pierini}, D., {Leech}, K.~J., {Tuffs}, R.~J., \& {Volk}, H.~J. 1999, \mnras, 303, L29

\bibitem[{{Pierini} {et~al.}(2003){Pierini}, {Leech}, \& {V{\"o}lk}}]{Pierini_2003}
{Pierini}, D., {Leech}, K.~J., \& {V{\"o}lk}, H.~J. 2003, \aap, 397, 871

\bibitem[{{Pilbratt} {et~al.}(2010){Pilbratt}, {Riedinger}, {Passvogel}, {Crone}, {Doyle}, {Gageur}, {Heras}, {Jewell}, {Metcalfe}, {Ott}, \& {Schmidt}}]{Pilbratt_2010}
{Pilbratt}, G.~L., {Riedinger}, J.~R., {Passvogel}, T., {et~al.} 2010, \aap, 518, L1

\bibitem[{{Pineda} {et~al.}(2018){Pineda}, {Fischer}, {Kapala}, {Stutzki}, {Buchbender}, {Goldsmith}, {Ziebart}, {Glover}, {Klessen}, {Koda}, {Kramer}, {Mookerjea}, {Sandstrom}, {Scoville}, \& {Smith}}]{Pineda_2018}
{Pineda}, J.~L., {Fischer}, C., {Kapala}, M., {et~al.} 2018, \apjl, 869, L30

\bibitem[{{Poglitsch} {et~al.}(2010){Poglitsch}, {Waelkens}, {Geis}, {Feuchtgruber}, {Vandenbussche}, {Rodriguez}, {Krause}, {Renotte}, {van Hoof}, {Saraceno}, {Cepa}, {Kerschbaum}, {Agn{\`e}se}, {Ali}, {Altieri}, {Andreani}, {Augueres}, {Balog}, {Barl}, {Bauer}, {Belbachir}, {Benedettini}, {Billot}, {Boulade}, {Bischof}, {Blommaert}, {Callut}, {Cara}, {Cerulli}, {Cesarsky}, {Contursi}, {Creten}, {De Meester}, {Doublier}, {Doumayrou}, {Duband}, {Exter}, {Genzel}, {Gillis}, {Gr{\"o}zinger}, {Henning}, {Herreros}, {Huygen}, {Inguscio}, {Jakob}, {Jamar}, {Jean}, {de Jong}, {Katterloher}, {Kiss}, {Klaas}, {Lemke}, {Lutz}, {Madden}, {Marquet}, {Martignac}, {Mazy}, {Merken}, {Montfort}, {Morbidelli}, {M{\"u}ller}, {Nielbock}, {Okumura}, {Orfei}, {Ottensamer}, {Pezzuto}, {Popesso}, {Putzeys}, {Regibo}, {Reveret}, {Royer}, {Sauvage}, {Schreiber}, {Stegmaier}, {Schmitt}, {Schubert}, {Sturm}, {Thiel}, {Tofani}, {Vavrek}, {Wetzstein}, {Wieprecht}, \& {Wiezorrek}}]{Poglitsch_2010}
{Poglitsch}, A., {Waelkens}, C., {Geis}, N., {et~al.} 2010, \aap, 518, L2

\bibitem[{{Querejeta} {et~al.}(2021){Querejeta}, {Schinnerer}, {Meidt}, {Sun}, {Leroy}, {Emsellem}, {Klessen}, {Mu{\~n}oz-Mateos}, {Salo}, {Laurikainen}, {Be{\v{s}}li{\'c}}, {Blanc}, {Chevance}, {Dale}, {Eibensteiner}, {Faesi}, {Garc{\'\i}a-Rodr{\'\i}guez}, {Glover}, {Grasha}, {Henshaw}, {Herrera}, {Hughes}, {Kreckel}, {Kruijssen}, {Liu}, {Murphy}, {Pan}, {Pety}, {Razza}, {Rosolowsky}, {Saito}, {Schruba}, {Usero}, {Watkins}, \& {Williams}}]{querejeta_2020}
{Querejeta}, M., {Schinnerer}, E., {Meidt}, S., {et~al.} 2021, \aap, 656, A133

\bibitem[{{Ramambason} {et~al.}(2024){Ramambason}, {Lebouteiller}, {Madden}, {Galliano}, {Richardson}, {Saintonge}, {De Looze}, {Chevance}, {Abel}, {Hernandez}, \& {Braine}}]{Ramambason_2024}
{Ramambason}, L., {Lebouteiller}, V., {Madden}, S.~C., {et~al.} 2024, \aap, 681, A14

\bibitem[{{Rieke} {et~al.}(2004){Rieke}, {Young}, {Engelbracht}, {Kelly}, {Low}, {Haller}, {Beeman}, {Gordon}, {Stansberry}, {Misselt}, {Cadien}, {Morrison}, {Rivlis}, {Latter}, {Noriega-Crespo}, {Padgett}, {Stapelfeldt}, {Hines}, {Egami}, {Muzerolle}, {Alonso-Herrero}, {Blaylock}, {Dole}, {Hinz}, {Le Floc'h}, {Papovich}, {P{\'e}rez-Gonz{\'a}lez}, {Smith}, {Su}, {Bennett}, {Frayer}, {Henderson}, {Lu}, {Masci}, {Pesenson}, {Rebull}, {Rho}, {Keene}, {Stolovy}, {Wachter}, {Wheaton}, {Werner}, \& {Richards}}]{MIPS}
{Rieke}, G.~H., {Young}, E.~T., {Engelbracht}, C.~W., {et~al.} 2004, \apjs, 154, 25

\bibitem[{{Russell} {et~al.}(1980){Russell}, {Melnick}, {Gull}, \& {Harwit}}]{Russell_1980}
{Russell}, R.~W., {Melnick}, G., {Gull}, G.~E., \& {Harwit}, M. 1980, \apjl, 240, L99

\bibitem[{{Salo} {et~al.}(2015{\natexlab{a}}){Salo}, {Laurikainen}, {Laine}, {Comer{\'o}n}, {Gadotti}, {Buta}, {Sheth}, {Zaritsky}, {Ho}, {Knapen}, {Athanassoula}, {Bosma}, {Laine}, {Cisternas}, {Kim}, {Mu{\~n}oz-Mateos}, {Regan}, {Hinz}, {Gil de Paz}, {Menendez-Delmestre}, {Mizusawa}, {Erroz-Ferrer}, {Meidt}, \& {Querejeta}}]{Salo2015}
{Salo}, H., {Laurikainen}, E., {Laine}, J., {et~al.} 2015{\natexlab{a}}, \apjs, 219, 4

\bibitem[{{Salo} {et~al.}(2015{\natexlab{b}}){Salo}, {Laurikainen}, {Laine}, {Comer{\'o}n}, {Gadotti}, {Buta}, {Sheth}, {Zaritsky}, {Ho}, {Knapen}, {Athanassoula}, {Bosma}, {Laine}, {Cisternas}, {Kim}, {Mu{\~n}oz-Mateos}, {Regan}, {Hinz}, {Gil de Paz}, {Menendez-Delmestre}, {Mizusawa}, {Erroz-Ferrer}, {Meidt}, \& {Querejeta}}]{salo2015spitzer}
{Salo}, H., {Laurikainen}, E., {Laine}, J., {et~al.} 2015{\natexlab{b}}, \apjs, 219, 4

\bibitem[{{Sandstrom} {et~al.}(2013){Sandstrom}, {Leroy}, {Walter}, {Bolatto}, {Croxall}, {Draine}, {Wilson}, {Wolfire}, {Calzetti}, {Kennicutt}, {Aniano}, {Donovan Meyer}, {Usero}, {Bigiel}, {Brinks}, {de Blok}, {Crocker}, {Dale}, {Engelbracht}, {Galametz}, {Groves}, {Hunt}, {Koda}, {Kreckel}, {Linz}, {Meidt}, {Pellegrini}, {Rix}, {Roussel}, {Schinnerer}, {Schruba}, {Schuster}, {Skibba}, {van der Laan}, {Appleton}, {Armus}, {Brandl}, {Gordon}, {Hinz}, {Krause}, {Montiel}, {Sauvage}, {Schmiedeke}, {Smith}, \& {Vigroux}}]{Sandstrom_2013}
{Sandstrom}, K.~M., {Leroy}, A.~K., {Walter}, F., {et~al.} 2013, \apj, 777, 5

\bibitem[{{Sauty} {et~al.}(1998){Sauty}, {Gerin}, \& {Casoli}}]{Sauty}
{Sauty}, S., {Gerin}, M., \& {Casoli}, F. 1998, \aap, 339, 19

\bibitem[{{Schinnerer} {et~al.}(2006){Schinnerer}, {B{\"o}ker}, {Emsellem}, \& {Lisenfeld}}]{Schinnerer_2006}
{Schinnerer}, E., {B{\"o}ker}, T., {Emsellem}, E., \& {Lisenfeld}, U. 2006, \apj, 649, 181

\bibitem[{{Smith} {et~al.}(2017){Smith}, {Croxall}, {Draine}, {De Looze}, {Sandstrom}, {Armus}, {Beir{\~a}o}, {Bolatto}, {Boquien}, {Brandl}, {Crocker}, {Dale}, {Galametz}, {Groves}, {Helou}, {Herrera-Camus}, {Hunt}, {Kennicutt}, {Walter}, \& {Wolfire}}]{Smith_2017}
{Smith}, J.~D.~T., {Croxall}, K., {Draine}, B., {et~al.} 2017, \apj, 834, 5

\bibitem[{{Smith} {et~al.}(2019){Smith}, {Clark}, {De Looze}, {Lamperti}, {Saintonge}, {Wilson}, {Accurso}, {Brinks}, {Bureau}, {Chung}, {Cigan}, {Clements}, {Dharmawardena}, {Fanciullo}, {Gao}, {Gao}, {Gear}, {Gomez}, {Greenslade}, {Hwang}, {Kemper}, {Lee}, {Li}, {Lin}, {Liu}, {Moln{\'a}r}, {Mok}, {Pan}, {Sargent}, {Scicluna}, {Smith}, {Urquhart}, {Williams}, {Xiao}, {Yang}, \& {Zhu}}]{Smith_2019}
{Smith}, M. W.~L., {Clark}, C. J.~R., {De Looze}, I., {et~al.} 2019, \mnras, 486, 4166

\bibitem[{{Stacey} {et~al.}(1991){Stacey}, {Geis}, {Genzel}, {Lugten}, {Poglitsch}, {Sternberg}, \& {Townes}}]{Stacey_1991}
{Stacey}, G.~J., {Geis}, N., {Genzel}, R., {et~al.} 1991, \apj, 373, 423

\bibitem[{{Stacey} {et~al.}(2010){Stacey}, {Hailey-Dunsheath}, {Ferkinhoff}, {Nikola}, {Parshley}, {Benford}, {Staguhn}, \& {Fiolet}}]{stacey_2010}
{Stacey}, G.~J., {Hailey-Dunsheath}, S., {Ferkinhoff}, C., {et~al.} 2010, \apj, 724, 957

\bibitem[{{Stewart} \& {Quijada}(2000)}]{IRAC}
{Stewart}, K.~P. \& {Quijada}, M.~A. 2000, in Society of Photo-Optical Instrumentation Engineers (SPIE) Conference Series, Vol. 4131, Infrared Spaceborne Remote Sensing VIII, ed. M.~{Strojnik} \& B.~F. {Andresen}, 218--227

\bibitem[{{Sutter} {et~al.}(2019){Sutter}, {Dale}, {Croxall}, {Pelligrini}, {Smith}, {Appleton}, {Beir{\~a}o}, {Bolatto}, {Calzetti}, {Crocker}, {De Looze}, {Draine}, {Galametz}, {Groves}, {Helou}, {Herrera-Camus}, {Hunt}, {Kennicutt}, {Roussel}, \& {Wolfire}}]{Sutter}
{Sutter}, J., {Dale}, D.~A., {Croxall}, K.~V., {et~al.} 2019, \apj, 886, 60

\bibitem[{{Sutter} \& {Fadda}(2022)}]{Sutter_2022}
{Sutter}, J. \& {Fadda}, D. 2022, \apj, 926, 82

\bibitem[{{Tielens} \& {Hollenbach}(1985)}]{Tielens_1985}
{Tielens}, A.~G.~G.~M. \& {Hollenbach}, D. 1985, \icarus, 61, 40

\bibitem[{{Tran} {et~al.}(2023){Tran}, {Williams}, {Levesque}, {Lazzarini}, {Dalcanton}, {Dolphin}, {Koplitz}, {Smercina}, \& {Telford}}]{Tran_2023}
{Tran}, D., {Williams}, B., {Levesque}, E., {et~al.} 2023, \apj, 954, 211

\bibitem[{{Vacca} {et~al.}(2020){Vacca}, {Clarke}, {Perera}, {Fadda}, \& {Holt}}]{pipeline}
{Vacca}, W., {Clarke}, M., {Perera}, D., {Fadda}, D., \& {Holt}, J. 2020, in Astronomical Society of the Pacific Conference Series, Vol. 527, Astronomical Data Analysis Software and Systems XXIX, ed. R.~{Pizzo}, E.~R. {Deul}, J.~D. {Mol}, J.~{de Plaa}, \& H.~{Verkouter}, 547

\bibitem[{{Vallini} {et~al.}(2015){Vallini}, {Gallerani}, {Ferrara}, {Pallottini}, \& {Yue}}]{Vallini_2015}
{Vallini}, L., {Gallerani}, S., {Ferrara}, A., {Pallottini}, A., \& {Yue}, B. 2015, \apj, 813, 36

\bibitem[{{Walter} {et~al.}(2008){Walter}, {Brinks}, {de Blok}, {Bigiel}, {Kennicutt}, {Thornley}, \& {Leroy}}]{Walter_2008}
{Walter}, F., {Brinks}, E., {de Blok}, W.~J.~G., {et~al.} 2008, \aj, 136, 2563

\bibitem[{{Walter} {et~al.}(2012){Walter}, {Decarli}, {Carilli}, {Riechers}, {Bertoldi}, {Wei{\ss}}, {Cox}, {Neri}, {Maiolino}, {Ouchi}, {Egami}, \& {Nakanishi}}]{Walter}
{Walter}, F., {Decarli}, R., {Carilli}, C., {et~al.} 2012, \apj, 752, 93

\bibitem[{{Weingartner} \& {Draine}(2001)}]{Weingartner_2001}
{Weingartner}, J.~C. \& {Draine}, B.~T. 2001, \apj, 548, 296

\bibitem[{{Wolfire} {et~al.}(2010){Wolfire}, {Hollenbach}, \& {McKee}}]{Wolfire_2010}
{Wolfire}, M.~G., {Hollenbach}, D., \& {McKee}, C.~F. 2010, \apj, 716, 1191

\bibitem[{{Wolfire} {et~al.}(1995){Wolfire}, {Hollenbach}, {McKee}, {Tielens}, \& {Bakes}}]{Wolfire_1995}
{Wolfire}, M.~G., {Hollenbach}, D., {McKee}, C.~F., {Tielens}, A.~G.~G.~M., \& {Bakes}, E.~L.~O. 1995, \apj, 443, 152

\bibitem[{{Wolfire} {et~al.}(2003){Wolfire}, {McKee}, {Hollenbach}, \& {Tielens}}]{Wolfire_2003}
{Wolfire}, M.~G., {McKee}, C.~F., {Hollenbach}, D., \& {Tielens}, A.~G.~G.~M. 2003, \apj, 587, 278

\bibitem[{{Wright} {et~al.}(2010){Wright}, {Eisenhardt}, {Mainzer}, {Ressler}, {Cutri}, {Jarrett}, {Kirkpatrick}, {Padgett}, {McMillan}, {Skrutskie}, {Stanford}, {Cohen}, {Walker}, {Mather}, {Leisawitz}, {Gautier}, {McLean}, {Benford}, {Lonsdale}, {Blain}, {Mendez}, {Irace}, {Duval}, {Liu}, {Royer}, {Heinrichsen}, {Howard}, {Shannon}, {Kendall}, {Walsh}, {Larsen}, {Cardon}, {Schick}, {Schwalm}, {Abid}, {Fabinsky}, {Naes}, \& {Tsai}}]{wise}
{Wright}, E.~L., {Eisenhardt}, P. R.~M., {Mainzer}, A.~K., {et~al.} 2010, \aj, 140, 1868

\bibitem[{{Young} {et~al.}(2012){Young}, {Becklin}, {Marcum}, {Roellig}, {De Buizer}, {Herter}, {G{\"u}sten}, {Dunham}, {Temi}, {Andersson}, {Backman}, {Burgdorf}, {Caroff}, {Casey}, {Davidson}, {Erickson}, {Gehrz}, {Harper}, {Harvey}, {Helton}, {Horner}, {Howard}, {Klein}, {Krabbe}, {McLean}, {Meyer}, {Miles}, {Morris}, {Reach}, {Rho}, {Richter}, {Roeser}, {Sandell}, {Sankrit}, {Savage}, {Smith}, {Shuping}, {Vacca}, {Vaillancourt}, {Wolf}, \& {Zinnecker}}]{Young_2012}
{Young}, E.~T., {Becklin}, E.~E., {Marcum}, P.~M., {et~al.} 2012, \apjl, 749, L17

\bibitem[{{Zanella} {et~al.}(2018){Zanella}, {Daddi}, {Magdis}, {Diaz Santos}, {Cormier}, {Liu}, {Cibinel}, {Gobat}, {Dickinson}, {Sargent}, {Popping}, {Madden}, {Bethermin}, {Hughes}, {Valentino}, {Rujopakarn}, {Pannella}, {Bournaud}, {Walter}, {Wang}, {Elbaz}, \& {Coogan}}]{Zanella_2018}
{Zanella}, A., {Daddi}, E., {Magdis}, G., {et~al.} 2018, \mnras, 481, 1976

\bibitem[{{Zhang} {et~al.}(1993){Zhang}, {Wright}, \& {Alexander}}]{zhang_1993}
{Zhang}, X., {Wright}, M., \& {Alexander}, P. 1993, \apj, 418, 100

\end{thebibliography}

\appendix
 \section{Comparison of FIFI-LS \CII\ and PACS}
 \label{appendix-crossc}

To compare the flux calibration of our 3 maps we use the KINGFISH \CII\ PACS maps from the \textit{Herschel} KINGFISH survey \citep{Kennicutt_2011}, using the calibration block of the data \footnote{\url{https://irsa.ipac.caltech.edu/data/Herschel/KINGFISH/index.html}}.

\begin{figure*}
    \centering
    \includegraphics[width=\textwidth]{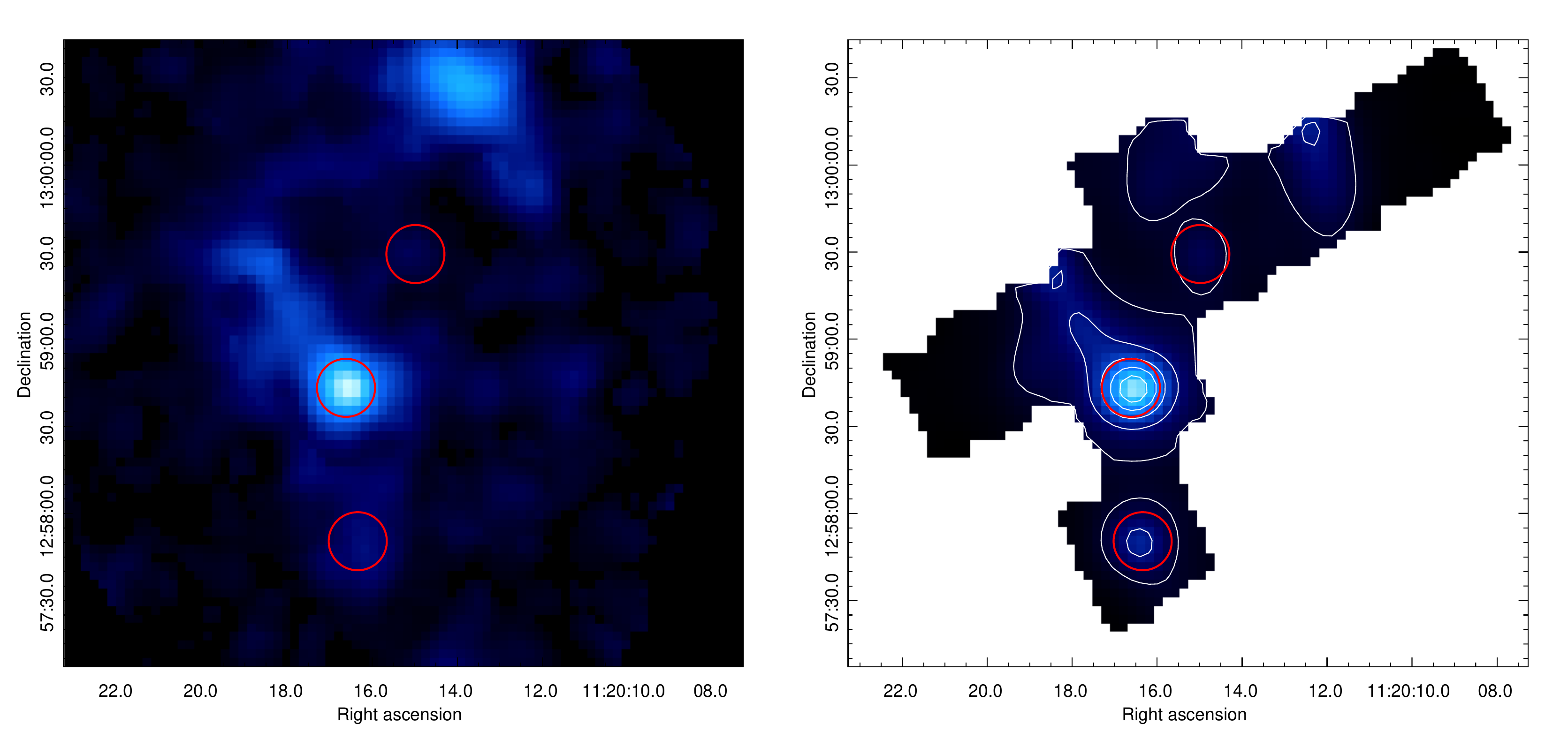}
    \caption{Comparison of PACS and FIFI-LS fluxes for  NGC\,3627. Three $20\arcsec$ diameter apertures (red), which are point-like sources without edge effects in the PACS coverage, were chosen to compare with the FIFI-LS data. Flux deviations towards these positions are 7.5\%, 21\%, and 10\%,  within the 20\% calibration uncertainty.}
    \label{fig:enter-label}
\end{figure*}

Within the \CII\ fields available from PACS we have chosen a total of 8 apertures containing bright point-like \CII\ sources clear of the edges of the PACS fields to avoid edge effects (see example for NGC\,3627 in Fig.\,\ref{fig:enter-label}). We used apertures with relatively small $20\arcsec$ diameters to also check for any potential issues with pointing or resolution as well. The PACS data was smoothed to match the $15.6\arcsec$ resolution of the FIFI-LS data. In all 8 apertures the deviation between FIFI-LS and PACS is less than 21\%. Using a conservatively small 10\% flux calibration error for PACS this shows that we can assume a flux uncertainty of 20\% for our FIFI-LS data.

\begin{table*}
        \caption{Calibration comparison values.}
        \centering

\begin{tabular}{l r r r r r r}
\hline \hline
NGC\,3627 	&RA 	&Dec 	&radius (\arcsec) 	&PACS W/m$^{2}$ 	&FIFI-LS w/m$^{2}$ &	deviation (\%)\\
1&  	11:20:14.9845 	& +12:59:29.337 	& 10 	& 3.54E-16 	& 3.80E-16 & 	7.52E-02\\
2 & 	11:20:16.6189 	& +12:58:43.241 	& 10 	& 1.69E-15 & 	2.04E-15 	& 2.05E-01\\
3 	& 11:20:16.3427 	& +12:57:50.415 	& 10 	& 5.82E-16&  	6.42E-16 	& 1.03E-01\\

\hline
NGC\,4321	&& 	& 	&	& &	\\
1 	&12:22:54.8258& 	+15:49:18.665 &	10 &	1.14E-15 &	1.17E-15 	&2.60E-02\\
2 &	12:22:59.1771 	&+15:48:54.260 	&10 	&3.44E-16 	&3.97E-16 	&1.55E-01\\
\hline
NGC\,6946	&& 	& 	&	& &	\\     
1& 	20:34:32.1186 &	+60:10:18.194 	&10 	&7.07E-16 	&5.80E-16 &	1.81E-01\\
2 &	20:34:52.4067 	&+60:09:13.314 &	10 &	2.85E-15 &	2.71E-15 	&4.92E-02\\
3 &	20:35:11.5149 	&+60:08:56.475 &	10 &	1.06E-15 	&8.94E-16 &	1.54E-01   \\ 
\hline
  && 	& 	&	&Mean deviation: 	&1,18E-01\\   
\hline\hline
\end{tabular}
\end{table*}

\section{Effect of $\alpha_\text{CO}$ on the relation between \CII\ and molecular gas mass  }

 \begin{figure*}
    \centering
    \includegraphics[width=\textwidth]{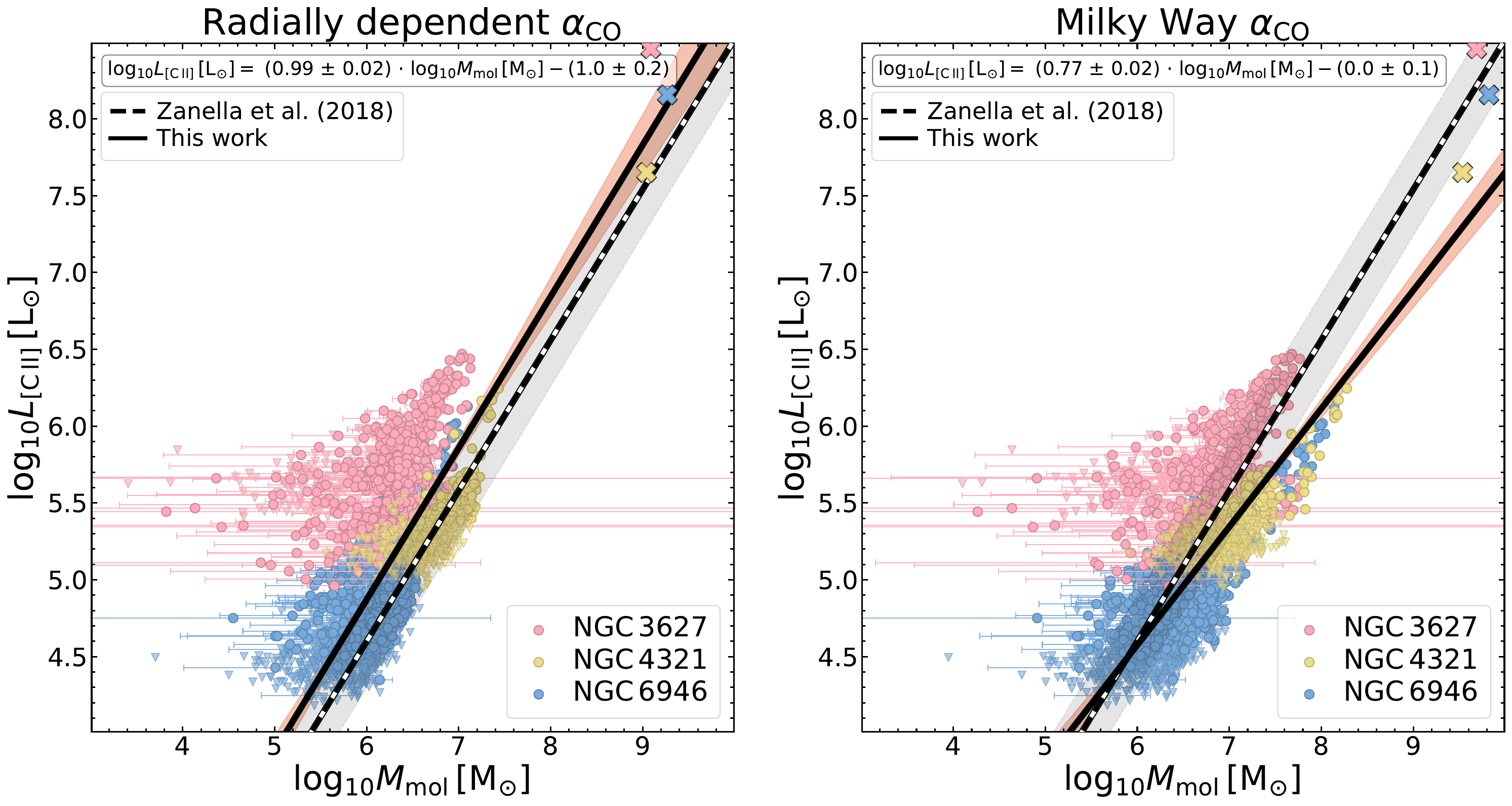}
    \caption{Similar to Fig.\,\ref{fig:zanella}. Left: Using the radially dependent $\alpha_\text{CO}$ from \citet{Sandstrom_2013}. Right: Using the constant Milky Way $\alpha_\text{CO} = 4.4 {\rm\, M_{\odot}\,pc^{-2}\,\left(K\,km\,s^{-1}\right)^{-1}}$.}
    \label{fig:alpha_co_comp}
\end{figure*}

One of the more striking differences between the SOFIA FIFI-LS observations presented in our work and the previous work done by \citet{Zanella_2018} is in the deviation of the fit for the galaxy NGC\,3627, and, to a lesser extent, NGC\,6946 and NGC\,4321, as shown in Fig.\,\ref{fig:zanella}. In \citet{Zanella_2018}, the authors model their molecular mass budget using a constant Milky Way value of the CO-to-H$_2$ conversion factor:
\begin{equation}
    \alpha_\text{CO} = 4.4 {\rm \, M_{\odot} \, pc^{-2} \, \left(K\,km\,s^{-1}\right)^{-1}}\quad .
\end{equation}

However, in our study, we used the radially dependent values from \citet{Sandstrom_2013} (see Fig. 7 of their work). All 3 galaxies from our sample have been studied in \citet{Sandstrom_2013} and have their radially dependent conversion factors accurately constrained. It is normally not recommended to use the Milky Way CO-to-H$_2$ conversion factor for resolved galaxies since the values for $\alpha_\text{CO}$ can vary significantly, depending on the local conditions. For that reason we choose against using the Milky Way values in our analysis.

To demonstrate the effect that the choice of the $\alpha_\text{CO}$ has on the relationship between \CII\ and molecular gas mass, we plot our entire dataset using a constant Milky Way $\alpha_\text{CO}$ value of 4.4 and compare this to our results from Section \ref{sec:mol_gas_tracer} using the spatially varying $\alpha_\text{CO}$ values (Fig.\,\ref{fig:alpha_co_comp}.) While the constant Milky Way $\alpha_\text{CO}$ value shifts the data for the three galaxies to higher molecular gas masses, there are still wide deviations of the \CII-$\alpha_\text{CO}$ relation compared to that of \citet{Zanella_2018} from studying unresolved galaxies. 

One of the main conclusions of our work is that there is not one picture of \CII\ as a reliable tracer of star formation throughout galaxies, but, as spatial studies like this show, wide variations exist due to local conditions, especially when mapping over full galaxies. Wide differences in \CII\ vs molecular gas mass are seen in these figures even within individual galaxies. This demonstrates the importance of studying full galaxies with high spatial resolution and applying accurate values of $\alpha_\text{CO}$.
\label{appendixb}
\end{document}